\documentclass[fleqn,usenatbib]{mnras}
\usepackage{rotating} 
\usepackage{booktabs} 

\usepackage{float}

\usepackage{newtxtext}
\usepackage[slantedGreek]{newtxmath}

\usepackage{placeins}
\usepackage{newtxtext,newtxmath}
\usepackage{subcaption}
\usepackage{booktabs}
\usepackage{lastpage}
\usepackage{caption}

\usepackage[T1]{fontenc}

\def\h {^{\mathrm{h}}}
\def\m {^{\mathrm{m}}}
\def\deg {^{\circ} }
\def\sqdeg {~deg$^2$}

\def\arcmin{\hbox{$^\prime$}}
\def\arcsec{\hbox{$^{\prime\prime}$}}

\def\fs{\hbox{$.\!\!^{\rm s}$}}

\def\farcs{\hbox{$.\!\!^{\prime\prime}$}}

\DeclareRobustCommand{\VAN}[3]{#2}
\let\VANthebibliography\thebibliography
\def\thebibliography{\DeclareRobustCommand{\VAN}[3]{##3}\VANthebibliography}

\usepackage{graphicx}	
\usepackage{amsmath}	

\title[Dynamical modelling of GRQs]{Dynamical modelling of giant radio quasars in the HETDEX Spring Field}

\author[Franzen et al.]{T.~M.~O.~Franzen$^{1,2}$\thanks{E-mail: thomas.franzen@skao.int},
J.~Machalski$^{3}$,
M.~Jamrozy$^{3}$,
R.~Morganti$^{2,4}$,
D.~A.~Green$^{5}$,
Y.~C.~Perrott$^{6}$
\newauthor
and T.~W.~Shimwell$^{2,7}$
\\
$^{1}$SKA Observatory, Jodrell Bank, Lower Withington, Macclesfield SK11 9FT, UK\\
$^{2}$ASTRON, the Netherlands Institute for Radio Astronomy, Oude Hoogeveensedijk 4, 7991 PD Dwingeloo, The Netherlands\\
$^{3}$Astronomical Observatory, Jagiellonian University, ul. Orla 171, PL-30244 Krakow, Poland\\
$^{4}$Kapteyn Astronomical Institute, University of Groningen, Postbus 800, 9700 AV Groningen, The Netherlands\\
$^{5}$Astrophysics Group, Cavendish Laboratory, JJ Thomson Avenue, Cambridge CB3 0US, UK\\
$^{6}$School of Chemical and Physical Sciences, Victoria University of Wellington, PO Box 600, Wellington 6140, New Zealand\\
$^{7}$Leiden Observatory, Leiden University, PO Box 9513, 2300 RA Leiden, The Netherlands
}

\date{Accepted XXX. Received YYY; in original form ZZZ}

\pubyear{2022}

\begin{document}
\label{firstpage}
\pagerange{\pageref{firstpage}--\pageref{lastpage}}
\maketitle

\begin{abstract}
Giant radio sources are defined as extragalactic radio sources, hosted by galaxies or quasars, with linear sizes $\geq 0.7$~Mpc. They are thought to represent the final stage of the evolution of radio galaxies whose sizes range from pc to Mpc scales. We analyse in detail the radio morphology and spectra between 54~MHz and 4.85/15.5~GHz of 15 giant radio quasars (GRQs) in the HETDEX Spring Field, and fit dynamical evolution models to the sources' observational properties to study their physical parameters (age of the lobes, jet power, ambient medium density etc.). We compare the physical parameters of the GRQs with published results for a compiled sample of Fanaroff--Riley type II radio sources. We find that the GRQs evolve in significantly lower medium densities, both at the central core radius and in front of the lobes, than smaller-sized radio quasars (RQs) with similar jet powers. The derived central core densities for both populations are, however, highly sensitive to the assumed ambient medium density profile. For both populations combined, the jet power, $Q_{\rm j}$, is anti-correlated with the age of the lobes, $t_{\ell}$ ($R = -0.67$, where $R$ is the Pearson correlation coefficient), as well as the linear size, $D_{\ell}$ ($R=-0.32$), but $t_{\ell}$ is a much stronger indicator of $Q_{\rm j}$ than $D_{\ell}$. Using a Spearman partial rank correlation analysis, we demonstrate that there is a fundamental relation between $Q_{\rm j}$, $t_{\ell}$, and $D_{\ell}$, despite the strong underlying correlation between age and size.
\end{abstract}



\section{Introduction}\label{Introduction}

Several analytical models have been developed to describe the dynamic and energetic properties of powerful Fanaroff--Riley type II \citep[FRII;][]{fanaroff1974} radio sources \citep[e.g.][]{begelman1989,kaiser1997b,blundell1999,machalski2007,hardcastle2018,turner2023}. According to these models, the galaxy's active galactic nucleus (AGN) emits energy in the form of jets composed of relativistic particles, channeling their way through the external environment. The jets slow down upon impact with dense material, dissipating their energy through shocks, resulting in the formation of bright regions at the ends of the jets, known as hotspots. The relativistic particles flow out from the hotspots to form extended radio synchrotron-emitting lobes that dominate the radiative output of these systems at radio frequencies. The spectrum of the radio galaxy (RG) steepens over time as the synchrotron particles in the lobes lose energy through synchrotron and inverse Compton (IC) processes.

Detailed modelling of the observed radio emission of FRIIs can therefore be used to derive their physical parameters, such as the age of the lobes, jet power and density of the ambient medium into which the radio structure evolves. In particular, \cite{machalski2021} published dynamical models for 361 FRII-type radio sources, selected from the 5C6/5C7 \citep{pearson1978}, 3CRR \citep{laing1983}, Bologna B2 \citep{deruiter1986,fanti1986,fanti1987}, Green Bank GB/GB2 \citep{machalski1998} and 6CE \citep{rawlings2001} samples, hereafter referred to as the Atlas. Most of these sources are identified as RGs, with only a minority (22 per cent) classified spectroscopically as quasars.

According to the unified model of AGN \citep{antonucci1993,urry1995}, the distinction between a RG and a radio-loud quasar primarily relates to the observer's viewing angle relative to the AGN. Radio-loud quasars are viewed at a small angle ($< 45$~deg) to the jet axis, allowing a clear view of the central AGN and the broad emission line region. RGs are viewed at a large angle ($> 45$~deg) to the jet axis, where a dusty torus blocks the direct view of the nucleus, leaving only the narrow emission lines visible, or none at all. Radio-loud quasars are almost exclusively associated with high-excitation radio galaxies (HERGs), fuelled at high rates through radiatively efficient classic accretion discs. In contrast, RGs can be associated with either HERGs or low-excitation radio galaxies (LERGs), powered at significantly lower rates by radiatively inefficient accretion flows \citep{best2012}.

The Atlas includes six giant radio quasars (GRQs) with deprojected linear sizes, $D_{\ell}$, greater than or equal to 0.7~Mpc. Giant radio sources (GRSs) are thought to represent the final stage of the evolution of RGs whose sizes range from pc to Mpc scales. They are much rarer than smaller-sized RGs and the exact reasons for their gigantic sizes are still debated. A number of explanations have been proposed, including the following:
\begin{enumerate}
    \item GRSs have jets that are much more powerful than normal RGs, which allow them to overcome the resistance of the Intergalactic Medium (IGM) to their large-scale growth \citep[e.g.][]{wiita1989,gopal-krishna1989}.
    \item GRSs are very old and have had enough time to expand to large distances \citep[e.g.][]{subrahmanyan1996,jamrozy2008}.
    \item The growth of GRSs to large linear sizes is favoured in low-density environments where the jets can propagate freely without significant interference \citep[e.g.][]{mack1998,schoenmakers2000,stuardi2020}.
\end{enumerate}
The above hypotheses are not mutually exclusive; all three could be at play at the same time.

Since the lobes of GRSs extend far beyond the interstellar medium of the host galaxy, they act as sensitive probes of the Warm-Hot IGM \citep[WHIM;][]{dave2001}, tracing the magnetic fields and gas densities of the cosmic web's filaments \citep[e.g.][]{osullivan2019,oei2022}. GRQs are exceptionally bright at both optical and radio wavelengths, allowing them to be seen at much higher redshifts than giant radio galaxies (GRGs). This makes them excellent tools for mapping the evolution of the cosmic web in the early Universe.

In this paper, we present dynamical evolution models for 15 GRQs, most of which were discovered by \cite{dabhade2020} in the Hobby--Eberly Telescope Dark Energy Experiment (HETDEX) Spring Field \citep[right ascension $10\h 45\m$ to $15\h 30\m$ and declination $45\deg$ to $57\deg$;][]{hill2008}. Using the DYNAGE algorithm of \cite{machalski2007}, we model the radio spectra of the GRQs between 54~MHz and 4.85/15.5~GHz to estimate their physical parameters. This numerical code, which was also applied to the Atlas by \cite{machalski2021}, is based on the \cite{kaiser1997} analytical model for the dynamical evolution of classical doubles. We examine how the ambient medium densities, jet powers and ages of the GRQs compare with those of the smaller-sized radio quasars (RQs) in Atlas to improve our understanding of what drives the exceptional growth of GRQs.

Using a preliminary version of the Atlas, \cite{wojtowicz2021} noticed a positive and statistically significant correlation between the integrated spectral index of the non-thermal emission from the radio lobes and the ambient medium density surrounding the sources. They proposed that the discovered correlation could be used as a cosmological tool to estimate the ambient medium density for large samples of distant radio sources. Here, we test whether the correlations determined for the Atlas RGs and RQs separately are significantly different. We also investigate whether the correlation for the RQs still holds for our sample of GRQs, noting that only six of the 80 Atlas RQs have $D_{\ell} \geq 0.7$~Mpc.

Throughout this paper, we assume the convention for spectral index, $\alpha$, where $S \propto \nu^{-\alpha}$. The linear sizes and luminosities of the sources in our sample are calculated with a Hubble constant of $H_{0} = 71~\mathrm{km}~\mathrm{s}^{-1}~\mathrm{Mpc}^{-1}$, and matter and cosmological constant density parameters of $\Omega_{M} = 0.27$ and $\Omega_{\Lambda} = 0.73$, as adopted in the Atlas. This is different to the cosmology used by \cite{dabhade2020} ($H_{0} = 67.8~\mathrm{km}~\mathrm{s}^{-1}~\mathrm{Mpc}^{-1}$, $\Omega_{M} = 0.308$, $\Omega_{\Lambda} = 0.692$), from which the GRQs studied in this paper are selected.


\section{Selection of the GRQ sample}\label{Selection of the GRQ sample}

We select known GRQs in the HETDEX Spring Field, which is covered by the LOFAR Two-metre Sky Survey \citep[LoTSS;][]{shimwell2019,shimwell2022} at 144~MHz. LoTSS provides the combination of high surface brightness sensitivity and high spatial resolution (6~arcsec), which are key requirements in studying GRSs, especially at high redshifts. \cite{dabhade2020} identified 40 GRQs in this region of sky using LoTSS DR1 \citep{shimwell2019}. Of these 40 sources, 17 have sufficiently well-characterised radio spectra\footnote{The determination of the radio spectra of the GRQs is presented in Section~\ref{Radio spectra}. We require lobe flux densities measured at least at five observing frequencies with the lowest frequency equal to or less than 74~MHz and the highest frequency equal to or greater than 1.4~GHz.} to allow reliable parameters to be extracted from the dynamical modelling (see Section~\ref{Dynamical modelling}).

The largest angular sizes (LASs) of these sources are re-measured for the dynamical modelling and discrepancies of up to 36 per cent are found between our measurements and those in \cite{dabhade2020}. In \citeauthor{dabhade2020}, the LAS was taken as the angular separation between the points separated the furthest from one another on the $3\sigma$ contours of the source. Here, the LAS is taken as the angular separation between the brightest regions of the lobes or hotspots, which is more robust as the brightest regions are most likely the jet termination points. As a result, our LAS measurements are on average 8 per cent lower than those of \citeauthor{dabhade2020}. Four of the sources are found to have $D_{\ell} < 0.7$~Mpc and are removed from our sample. The inclination angles\footnote{The inclination angle is the angle between the jet axis and the line of sight ($i = 90$~deg means that the object lies in the plane of the sky).}, $i$, assumed in calculating $D_{\ell}$ are shown in Table~\ref{tab:GRG_properties} (see Section~\ref{Radio morphology} for a justification of these values).

\cite{amirkhanyan2016} found 50 GRQs across the entire sky using the NRAO Very Large Array (VLA) Sky Survey \citep[NVSS;][]{condon1998} and Faint Images of the Radio Sky at Twenty Centimetres \citep[FIRST;][]{becker1995} survey at 1.4~GHz, three of which lie in the HETDEX Spring Field and are not in the sample of \cite{dabhade2020}\footnote{These three sources are not in the sample of \citeauthor{dabhade2020} because they lie just outside the LoTSS DR1 mosaic, which does not fully cover the HETDEX Spring Field, as can be seen in figure~5 of \cite{shimwell2019}.}. Of these three sources, two (J1439+4550 and J1450+4549) have sufficient spectral data for the dynamical modelling and are added to our sample, resulting in a total of 15 GRQs whose observational parameters are summarised in Table~\ref{tab:GRG_properties}.

\begin{table*}
\caption{Observational data of the studied GRQs. The columns are: (1) source name; (2) RA of the optical QSO; (3) Dec of the optical QSO; (4) redshift; (5) largest angular size of the radio structure; (6) assumed inclination angle of the jet axis; (7) decimal logarithm of the total radio luminosity at 1.4~GHz; (8) deprojected linear size; (9) axial ratio. The deprojected linear size is given by $D_{\ell}' / \sin i$, where $D_{\ell}'$ is the projected linear size, as measured in the radio image. We note that $D_{\ell}$ and $AR$ determine the volume of the lobes, $V_{\ell}$.}
\label{tab:GRG_properties}
\begin{tabular}{ccccccccc}
\hline
Source name &  RA  &   Dec  & $z$ & LAS & $i$ & $\log_{10} P_{1.4}$ & $D_{\ell}$ & $AR$ \\
            & (h\;m\;s) &   ($\deg\;\arcmin\;\arcsec$)  &     & (arcmin)& (deg) & (W Hz$^{-1}$) & (Mpc) \\
(1)        &  (2)     &    (3)     & (4) &  (5)    &  (6)  & (7)   & (8)  &  (9)  \\
\hline
J1109+5104 & 11:09:35.39 & 51:04:02.3 & 1.179 & 1.77 & 90 & 26.37 & 0.88 & 5.0 \\
J1233+4902 & 12:33:05.44 & 49:02:51.9 & 1.352 & 1.62 & 90 & 26.02 & 0.82 & 3.0 \\
J1238+5325 & 12:38:07.76 & 53:25:55.9 & 0.347 & 2.73 & 90 & 25.43 & 0.80 & 1.8 \\
J1240+5334 & 12:40:12.46 & 53:34:37.2 & 0.293 & 2.73 & 90 & 25.85 & 0.71 & 3.2 \\
J1326+5358 & 13:26:46.56 & 53:58:17.4 & 0.410 & 2.38 & 70 & 25.35 & 0.83 & 4.1 \\
J1334+4813 & 13:34:18.63 & 48:13:17.1 & 2.208 & 1.47 & 70 & 27.31 & 0.78 & 3.3 \\
J1334+5501 & 13:34:11.70 & 55:01:24.9 & 1.245 & 1.57 & 90 & 27.58 & 0.79 & 2.3 \\
J1408+4738 & 14:08:32.49 & 47:38:37.4 & 1.436 & 1.28 & 70 & 26.69 & 0.70 & 3.0 \\
J1415+4909 & 14:15:54.37 & 49:09:21.2 & 1.371 & 1.77 & 90 & 26.34 & 0.90 & 3.7 \\
J1419+4837 & 14:19:35.98 & 48:37:43.2 & 0.496 & 2.07 & 90 & 25.91 & 0.75 & 3.2 \\
J1439+4550 & 14:39:32.67 & 45:50:28.3 & 1.836 & 1.70 & 90 & 26.58 & 0.87 & 2.7 \\
J1440+5026 & 14:40:13.98 & 50:26:01.9 & 1.574 & 2.18 & 70 & 26.39 & 1.19 & 3.9 \\
J1450+4549 & 14:50:38.83 & 45:49:54.6 & 1.622 & 1.47 & 90 & 27.18 & 0.80 & 5.0 \\
J1450+5300 & 14:50:57.28 & 53:00:07.8 & 0.918 & 3.03 & 90 & 26.83 & 1.43 & 4.3 \\
J1523+5203 & 15:23:11.07 & 52:03:03.5 & 0.517 & 2.38 & 90 & 25.47 & 0.89 & 2.1 \\
\hline
\end{tabular}
\end{table*}

\begin{figure}
\includegraphics[scale=0.50, angle=0, trim=0cm 0cm 0cm 0cm]{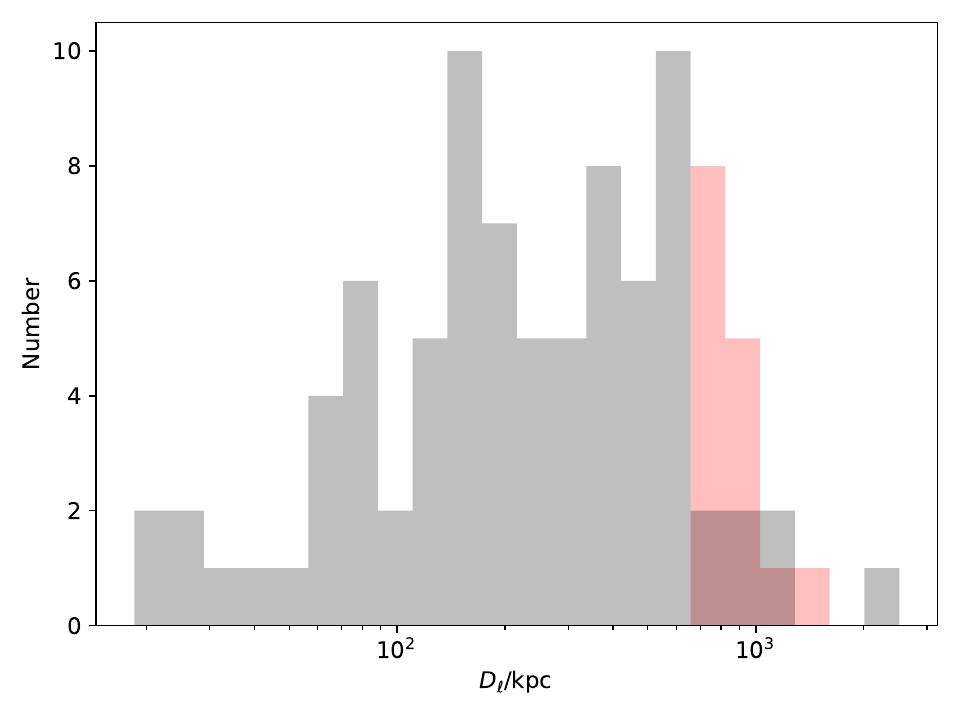}
\caption{Deprojected linear size distribution of the Atlas RQs (black) and our GRQs (red).}
\label{fig:linear_size_dist}
\end{figure}

\begin{figure}
\includegraphics[scale=0.55, angle=0, trim=0cm 0cm 0cm 0cm]{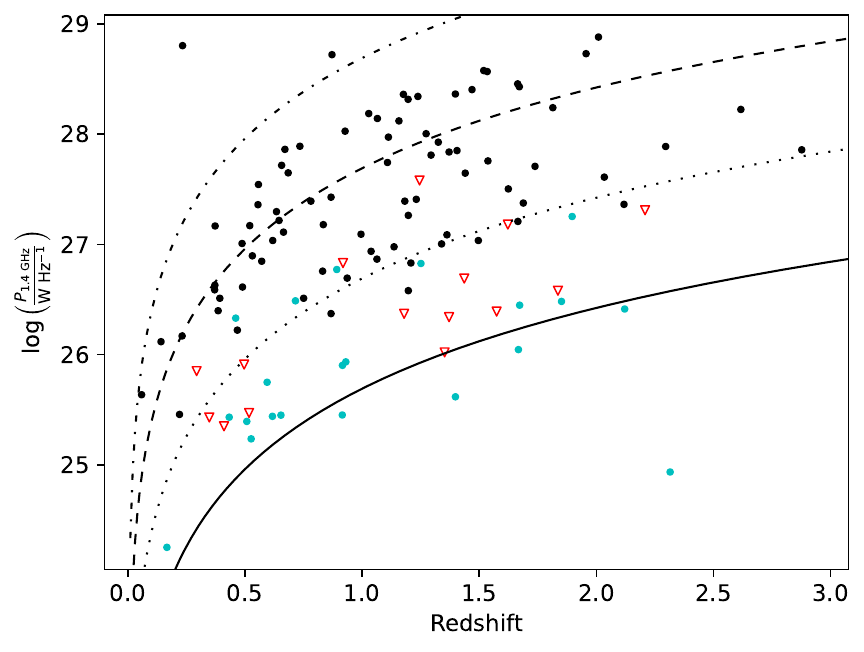}
\caption[short]{Total radio luminosity at 1.4~GHz versus redshift for the Atlas RQs (black dots), GRQs studied in this paper (red triangles) and remaining GRQs in \cite{dabhade2020} that do not have sufficiently well-characterised radio spectra to allow reliable parameters to be extracted from the dynamical modelling, for which we are able to derive reliable 1.4~GHz luminosities (cyan dots). In the Atlas, the 1.4~GHz flux densities used to derive the luminosities originate from NVSS. The redshifts are taken mostly from the NASA/IPAC database (NED) and the Sloan Digital Sky Survey sixteenth data release \citep[SDSS DR16;][]{ahumada2020}. In the \citeauthor{dabhade2020} sample, the 1.4~GHz flux densities are also derived from NVSS. Most of the redshifts originate from the LoTSS DR1 Value Added Catalogue \citep[VAC;][]{williams2019}; the redshift information in the LoTSS DR1 VAC was compiled mainly by using SDSS spectroscopic data. The solid, dotted, dashed and dot-dashed lines correspond to 1.4~GHz flux densities of 0.01, 0.1, 1.0 and 10~Jy, respectively.}
\label{fig:power_versus_redshift}
\end{figure}

Fig.~\ref{fig:linear_size_dist} compares the linear size distributions of the Atlas RQs and our GRQs. The Atlas RQs span a wide range in $D_{\ell}$ (19.5~kpc to 2.03~Mpc); the sample includes six GRQs with $D_{\ell} \geq 0.7$~Mpc. Fig.~\ref{fig:power_versus_redshift} shows the 1.4-GHz radio luminosity and redshift distributions of the two samples; the Atlas RQs are represented as black dots and the GRQs studied in this paper as red triangles. The mean 1.4-GHz radio luminosity of the Atlas RQs ($10^{27.45}~{\rm W}~{\rm Hz}^{-1}$) is about an order of magnitude higher than that of our GRQs ($10^{26.35}~{\rm W}~{\rm Hz}^{-1}$), but both samples probe 1.4-GHz radio luminosities down to $\sim 10^{25.5}~{\rm W}~{\rm Hz}^{-1}$. As mentioned in Section~\ref{Introduction}, the Atlas RQs are selected from a number of flux-limited samples, which together cover the entire sky north of declination $-30\deg$, and have $S_{1.4~{\rm GHz}} \gtrsim 100$~mJy. Our GRQs originate from LoTSS DR1, which covers a much smaller area of sky (424\sqdeg) and is much more sensitive than the surveys on which the Atlas sample is based. The LoTSS-selected GRQs have $S_{1.4~{\rm GHz}} \gtrsim 10$~mJy, while their mean redshift (1.12) is similar to that of the Atlas RQs (1.08). This explains why their 1.4~GHz luminosities are on average about an order of magnitude lower than those of the Atlas RQs.

The sample of \cite{dabhade2020} includes 27 GRQs which are not selected primarily because they do not have sufficiently well-characterised radio spectra to allow reliable parameters to be extracted from the dynamical modelling. The 1.4-GHz luminosity and redshift distributions of 21 of these sources with reliable spectral indices between 144--1400~MHz are represented as cyan dots in Fig.~\ref{fig:power_versus_redshift}. As expected, their luminosity distribution is similar to that of our GRQ sample. There appears to be a slight bias towards selecting GRQs from the sample of \citeauthor{dabhade2020} with higher luminosities at the same redshifts, which is not surprising since brighter sources will tend to have better-characterised radio spectra.

\section{Radio spectra}\label{Radio spectra}

In order to determine the radio spectra of our GRQs over a wide frequency range crucial for the dynamical modelling, we combine available radio survey data covering the HETDEX Spring Field including the LOFAR LBA Sky Survey first data release \citep[LoLSS DR1;][]{degasperin2023} at 54~MHz, VLA Low-frequency Sky Survey Redux \citep[VLSSr;][]{lane2012} at 74~MHz, LoTSS DR2 \citep{shimwell2022} at 144~MHz, Tata Institute for Fundamental Research GMRT Sky Survey First Alternative Data Release \citep[TGSS ADR1;][]{intema2017} at 148~MHz, Westerbork Northern Sky Survey \citep[WENSS;][]{rengelink1997} at 327~MHz, NVSS and FIRST at 1.4~GHz, VLA Sky Survey \citep[VLASS;][]{lacy2020} at 3~GHz, and Green Bank 6-cm \citep[GB6;][]{gregory1996} survey and its `New Catalogue' \citep{becker1991} at 4.85~GHz. We also observe four of the GRQs (J1109+5104, J1415+4909, J1439+4550 and J1450+4549), with the Arcminute Microkelvin Imager \citep[AMI;][]{zwart2008} to obtain a higher-frequency data point at 15.5~GHz. The sensitivity limits, angular resolutions, surface brightness limits and minimum \textit{uv} distances of these surveys and AMI-LA observations are shown in Table~\ref{tab:survey_properties}.

\begin{table*}
\caption{Characteristics of the surveys and AMI-LA observations from this paper used to measure the radio spectra of the GRQs.}
\label{tab:survey_properties}
\begin{tabular}{cccccc}
\hline
Survey/Telescope & Frequency & 5$\sigma$ sensitivity limit & Resolution & Surface brightness limit & Minimum \textit{uv} distance \\
            & (MHz) & (mJy beam$^{-1}$) & (arcsec)& (mJy arcmin$^{-2}$) & ($\lambda$) \\
\hline
LoLSS DR1 & 54 & 10 & 15 & 141 & 12.2 \\
VLSSr & 74 & 500 & 80 & 248 & 100 \\
LoTSS DR2 & 144 & 0.4 & 6 & 35 & 48 \\
TGSS ADR1 & 148 & 17.5 & 30 & 62 & 49 \\
WENSS & 327 & 18 & 70 & 11.7 & 59 \\
NVSS & 1400 & 2.5 & 45 & 3.9 & 163 \\
FIRST & 1400 & 0.65 & 5 & 83 & 980 \\
VLASS & 3000 & 0.6 & 2.5 & 310 & 2100  \\
GB6 & 4850 & 18 & 210 & 1.30 & 0 \\
AMI-LA & 15500 & $0.38-0.85$ & $\approx 40$ & $0.75-1.69$ & 930 \\
\hline
\end{tabular}
\end{table*}

Radio images of the 15 GRQs using LoLSS DR1, LoTSS DR2, VLASS and AMI data are shown in Fig.~\ref{fig:grq_overlays}. In this section, we describe the determination of the radio spectra of the GRQs.

\subsection{LoTSS DR2 data (144~MHz)}\label{LoTSS DR2 data}

The 144-MHz flux densities of the GRQs are derived from the high-resolution (6~arcsec) LoTSS DR2 mosaics, which have a typical rms noise of $\approx 80~\muup$Jy beam$^{-1}$. For each GRQ, we run the Aegean source finder \citep{hancock2012,hancock2018} on the LoTSS DR2 mosaic centred closest to the GRQ and use the island catalogue to derive its total flux density, $S_{\mathrm{tot}}$. The error on $S_{\mathrm{tot}}$ is given by 
\begin{equation}
\upDelta S_{\mathrm{tot}} = \sqrt { \left(\sigma_{\mathrm{rms}} \sqrt{A}\right)^2 + \left(\epsilon S_{\mathrm{tot}}\right)^2 } \, ,
\end{equation}
where $\sigma_{\mathrm{rms}}$ is the local rms noise, $A$ the total area covered by the source in units of the synthesised beam and $\epsilon$ the flux density calibration error. The systematic overall flux density scale error in LoTSS DR2 is estimated to be 10 per cent \citep{shimwell2022}; we therefore set $\epsilon$ to 0.1.

For 12 of the GRQs, a peak in the LoTSS DR2 mosaic is coincident with the host galaxy (i.e. the core is detected and not confused with the lobes/jets). For these GRQs, the core flux density, $S_{\mathrm{core}}$, is estimated from the Gaussian component catalogue, if possible. If the catalogue does not provide a reliable estimate of its flux density, we extract the peak flux density in a box with a width of a few pixels centred on the host galaxy. In the former case, the error on $S_{\mathrm{core}}$ is given by
\begin{equation}
\upDelta S_{\mathrm{core}} = \sqrt { \sigma_{\mathrm{fit}}^2 + \left(\epsilon S_{\mathrm{core}}\right)^2 } \, ,
\end{equation}
where $\sigma_{\mathrm{fit}}$ is the Gaussian fitting error on $S_{\mathrm{core}}$ which takes the local rms noise into account. In the latter case,
\begin{equation}
\upDelta S_{\mathrm{core}} = \sqrt { \sigma_{\mathrm{rms}}^2 + \left(\epsilon S_{\mathrm{core}}\right)^2 } \, .
\end{equation}
For the remaining GRQs, the core flux density is assumed to be negligible. The lobe flux density is derived by subtracting the core from the total flux density.

We note slight differences (up to 23 per cent) between the 144-MHz flux densities of the GRQs from LoTSS DR2 and those published in \cite{dabhade2020}, which were derived from LoTSS DR1. Both 144-MHz flux densities, weighted according to their respective errors, are used to determine the spectra of the modelled sources following the approach employed by \cite{machalski2021} in the Atlas when multiple flux densities are available at the same frequency.

\subsection{LoLSS DR1 data (54~MHz)}\label{LoLSS DR1 data}

The 54-MHz flux densities of the GRQs are derived from the LoLSS DR1 mosaic which has a typical rms noise of $1{-}2$ mJy beam$^{-1}$ and a resolution of 15~arcsec. We run Aegean on the LoLSS DR1 mosaic and measure the core (if possible) and lobe flux densities of the GRQs following the same method as described in Section~\ref{LoTSS DR2 data}. The core is detected and resolved in the LoLSS DR1 mosaic for only three of the GRQs.

The overall flux density scale error in LoLSS DR1 is estimated to be 6 per cent \citep{degasperin2023}. We take a conservative approach and set $\epsilon = 0.1$ when calculating the error on each flux density measurement.

\subsection{AMI data (15.5~GHz)}

\subsubsection{Observations}

AMI is located near Cambridge, UK, and consists of two separate telescopes, the Large Array (LA) and Small Array (SA). The GRQ data in this paper were collected in Dec 2019 and Jan 2020 using only the LA. The LA is an interferometer comprising eight 12.8-m-diameter, equatorially mounted dishes, with a range of baselines of $18{-}110$~m. It observes over the frequency band between $13{-}18$~GHz, with a central frequency of 15.5~GHz. It has a primary beam at 15.5~GHz of $\approx 5.5$~arcmin FWHM and a typical resolution of $\approx 30$~arcsec (this varies depending on the precise \textit{uv}-coverage of any observation). The telescope measures a single, linear polarisation (Stokes $I+Q$) and has a flux sensitivity of $\approx 3$ mJy beam$^{-1}$ for an integration time of 1~s.

J1109+5104, J1415+4909, J1439+4550 and J1450+4549 were observed using the upgraded correlator described in \cite{hickish2018}. The characteristics of the observations are shown in Table~\ref{tab:AMI_observations}. J1109+5104 and J1415+4909 were observed twice (at hour angles separated by $\approx 1$~h) for 1~h, and J1439+4550 and J1450+4549 were observed once for 1~h.

The other GRQs in our sample were not observed because more than two dishes were out of service, limiting the \textit{uv} coverage. The poor \textit{uv} coverage would have made it difficult to deconvolve the images and the angular resolution would likely have been too poor to separate the lobes from the core.

\begin{table*}
\centering
\caption{Characteristics of the AMI observations of the GRQs. The noise levels shown apply to the pointing centres and the beam position angles are measured from north to east. For J1109+5104 and J1415+4909, the noise levels, resolutions and beam position angles are for the combined data set.}
\label{tab:AMI_observations}
\begin{tabular}{@{} c c c c c c}
\hline
Source name & Observation date & Phase calibrator & Rms noise level & Resolution & Beam position angle \\
 & & & ($\muup$Jy beam$^{-1}$) & (arcsec$^{2}$) & (deg) \\
\hline
J1109+5104 & 25 Dec 2019, 1 Jan 2020 & J1041+5233 & 75 & $52 \times 28$ & 42 \\
J1415+4909 & 25, 31 Dec 2019 & J1417+4607 & 169 & $52 \times 28$ & 30 \\
J1439+4550 & 1 Jan 2020 & J1500+4751 & 140 & $60 \times 28$ & 16 \\
J1450+4549 & 31 Dec 2019 & J1500+4751 & 150 & $58 \times 27$ & 23 \\
\hline
\end{tabular}
\end{table*}

\subsubsection{Data reduction}

The data reduction is performed using CASA\footnote{https://casa.nrao.edu}. The full-frequency resolution data, consisting of 4096 1.22-MHz channels, are flagged for RFI using \textsc{rflag} before being averaged to 64 channels. Observations of the primary calibrator source 3C 286, close in time to the target observations, are used to calibrate the bandpass and set the flux density scale. The flux density scale of \cite{perley2013a} is used, but a correction for the fact that AMI measures $I+Q$ is applied using the integrated polarisation properties of 3C 286 from \cite{perley2013b} (this is an $\approx 4.5$ per cent correction). An amplitude correction for atmospheric amplitude variations is also applied using a noise-injection system, the `rain gauge' \citep[see][for more details]{zwart2008}. Interleaved observations of the secondary phase calibrators J1041+5233, J1417+4607 and J1500+4751 are used to correct the data for atmospheric and/or instrumental phase drift. Visibility weights are derived from the scatter of the visibilities using in-house software based on the \textsc{statwt} task.

The visibility data from the two separate observations of J1109+5104 and J1415+4909 are concatenated with \textsc{concat}. The data for all sources are then imaged over the whole bandwidth with \textsc{clean} using two Taylor coefficients to model the sky frequency dependence. Natural weighting is used to maximise the signal-to-noise ratio (SNR). A custom task is used to apply the LA primary beam correction. The final maps are shown by the magenta contours in Fig.~\ref{fig:grq_overlays}. The rms noise in these maps lies in the range $\approx 75{-}150~\muup$Jy beam$^{-1}$ and the beam size is $\approx 40$~arcsec.

We use the Owens Valley Radio Observatory (OVRO) blazar monitoring programme \citep{richards2011} as a cross-check of the AMI flux density calibration. In this monitoring programme, over 1000 sources were observed at 15~GHz at a typical cadence of twice a week between 2008 and 2020. The typical uncertainty on the OVRO flux densities is 3 per cent. The blazar sample includes two of the phase calibrators used in our AMI observations, J1041+5233 and J1500+4751. For each observing run containing J1041+5233 or J1500+4751 as the phase calibrator, we derive the 15-GHz flux density of the phase calibrator from a power-law fit to the 64 channel flux densities between 13 and 18~GHz. We then perform a flux density comparison using the closest measurement in time of the source in the OVRO blazar monitoring database.

We are able to find an OVRO measurement within three days of each of our observations. We find a $3{-}7$ per cent discrepancy between the AMI and OVRO flux densities. We note that the observed discrepancies could in part be due to flux density variability on timescales of a few days. We conclude that the flux density calibration of the telescope is accurate to within $<10$ per cent.

\subsubsection{Source extraction}\label{Source extraction}

We run Aegean on the AMI image of each GRQ and use the Gaussian component catalogue to derive the core and lobe flux densities. The error on each flux density measurement consists of a 10 per cent calibration error added in quadrature with the Gaussian fitting error.

For J1109+5104 and J1450+4549, the core and lobes are resolved into three components. For J1415+4909, a single point source coincident with the host galaxy is detected indicating that the emission primarily originates from the core. The AMI flux density is in good agreement with the value predicted by extrapolating the core flux densities from LoTSS DR2, NVSS and VLASS to 15.5~GHz assuming a linear power-law spectral energy distribution.

For J1439+4550, the northern lobe and hotspot are not detected, and higher resolution is needed to separate the southern lobe from the core. The FIRST and VLASS integrated flux densities of the northern hotspot are 4.61 and 3.47~mJy, respectively, which gives a spectral index of 0.37. Extrapolating these flux densities to 15.5~GHz gives a flux density of 1.88~mJy. However, the hotspot could be much fainter than 1~mJy at 15.5~GHz if it is a remnant hotspot and its spectrum steepens significantly between 3--15.5~GHz. The core flux density is estimated by extrapolating the core flux densities from LoTSS DR2, NVSS and VLASS to 15.5~GHz. The flux density of the southern lobe is derived by subtracting the core flux density from the total AMI flux density.

\subsection{Data at other radio frequencies}

The flux densities at other observed frequencies are determined similarly to those in the Atlas, particularly for the lobes, from values available for the entire source after subtracting the core flux density (if significant). The total flux densities are taken from survey catalogues including VLSSr, TGSS ADR1, WENSS, NVSS, FIRST, VLASS and GB6, and/or extragalactic radio source databases including \cite{kuehr1981} and the astrophysical CATalogs Support system (CATS\footnote{https://cats.sao.ru}). The core flux densities are taken mostly from the FIRST and VLASS catalogues.

\subsubsection*{Notes on the flux density measurements of individual sources}

\textbf{J1233+4902} \newline
In GB6, which has an angular resolution of $\approx 3$~arcmin, a strong confusing source lies about 1.6~arcmin to the south of J1233+4902. We measure the flux densities of the confusing source at 327, 1400 and 3000~MHz, and estimate its flux density at 4.85~GHz from the spectral fit. The flux density upper limit at 4.85~GHz results from the difference between the values of ($19 \pm 4$)~mJy in GB6 and 13.2~mJy determined from the spectral fit of the confusing source.

\noindent \textbf{J1439+4550} \newline
A point source is detected in the southern lobe at RA $14^{\rm h}39^{\rm m}33\fs91$, Dec $+45\deg50\arcmin00\arcsec$ in FIRST and VLASS (see Fig.~\ref{fig:grq_overlays}). It has a spectral index of $-0.1$ between 1.4 and 3~GHz and may be a weak confusing source. The lobe flux densities at 1.4 and 3~GHz are corrected for both this point source and the radio core.

\noindent \textbf{J1450+5300} \newline
The 74-MHz flux density of J1450+5300 is taken as the sum of the flux density of the western lobe from VLSS \citep{cohen2007}, measured at RA $14^{\rm h}50^{\rm m}47\fs38$, Dec $+53\deg00\arcmin00\farcs4$, and that of the eastern lobe from VLSSr.

\subsection{Results}

Table~\ref{tab:grq_flux_table} presents the observed lobe flux densities of the GRQs used to determine their spectra. For most sources, core flux densities are reliably measured at two or more frequencies if necessary and subtracted from the total flux densities to derive the lobe flux densities; the spectra are plotted separately for the core and lobes in Fig.~\ref{fig:grq_spectra}.

We acknowledge that the radio spectra are derived from observations with varying \textit{uv}-coverage, conducted using different instruments, configurations and frequency bands. While the total angular sizes of our targets range from $\approx 1.3-3.0$~arcmin, individual lobe sizes are smaller. At 15.5~GHz, the AMI-LA provides an LAS of $\approx 4$~arcmin. For the VLA L-band, the LAS is $\approx 2$~arcmin in the B-configuration and $\approx 15$~arcmin in the D-configuration, while for VLASS, it is $\approx 1$~arcmin. Consequently, some flux from the large-scale structures of the GRQs may be resolved out -- particularly in VLASS -- potentially steepening the measured spectral indices. However, the lobe spectra plotted in Fig.~\ref{fig:grq_spectra} show no systematic deficit at 3~GHz relative to the model fits (see Section~\ref{Dynamical modelling}). Based on previous work, we estimate that up to $\approx 10$ per cent of the flux density might be lost in the most extended sources, which remains within our assumed uncertainties. Notably, this study relies on integrated flux densities rather than a spatially resolved spectral index analysis.

\section{Radio morphology}\label{Radio morphology}

Most GRQs in our sample have quite symmetrical radio structures in the FIRST and VLASS images, as shown in Fig.~\ref{fig:grq_overlays}, indicating that their inclination angles are likely close to 90~deg. \cite{kuzmicz2012} investigated the radio and optical properties of 45 GRQs across the entire sky and most of these were also found to have large inclinations ($i > 60$~deg) based on their radio structures at 1.4~GHz. The fact that most of the GRQs have large inclinations is probably a selection effect resulting from the projected linear size cut.

The inclination angles assumed in the dynamical modelling are shown in Table~\ref{tab:GRG_properties}. For sources with considerable asymmetries in their radio lobe properties, $i$ is set to 70~deg. Otherwise, $i$ is assumed to be 90~deg. The influence of the value of $i$ on other model parameters is described in Section~\ref{Dynamical modelling}.

In SDSS DR18 \citep{almeida2023}, 14 of the GRQs are classified as Quasi-Stellar Objects (QSOs), 11 of which have broad emission lines. This result appears to be in conflict with AGN unification schemes \citep{urry1995}, where quasars have $i < 45$~deg. For objects with large inclinations, the broad-line region should be partially or totally hidden by a dusty torus. A possible explanation for the prominence of broad emission lines in the optical spectra of our GRQs, despite their large inclinations, is that some AGN have no dusty torus \citep{elitzur2008}, a clumpy or receding torus \citep{nenkova2008,honig2007}, or the outer accretion disc/torus is misaligned with respect to the black hole's spin axis \citep{bardeen1975,schmitt2003}, allowing broad emission lines to be observed even in quasars close to the plane of the sky.

\section{Dynamical modelling}\label{Dynamical modelling}

The analytical model for the evolution of FRII-type radio sources defined by \cite{kaiser1997b}, hereafter referred to as KDA, describes the synchrotron emission from the relativistic electrons within the radio lobes of a source, accounting for energy losses due to synchrotron radiation, IC scattering and adiabatic expansion. Assuming values of the model-free parameters, including the exponent of the power law of the electron energy distribution injected by the jets into the lobes, $p=1+2\alpha_{\rm inj}$, where $\alpha_{\rm inj}$ is the injection spectral index, the jet power, $Q_{\rm j}$, the lobes' lifetime, $t_{\ell}$, and the central density at the core radius, $\rho_{0}$, the KDA model predicts the lobes' expansion and radio luminosity evolution with time.

The dynamical modelling of our GRQ sample is performed using the DYNAGE algorithm \citep{machalski2007}, which is based on the KDA model. This numerical code solves the inverse problem, that is a determination of the above four free parameters by fitting the model to four observables: the linear extent, $D_{\ell}$, and volume, $V_{\ell}$, of the lobes, and the spectral indices (i.e. tangents to the spectrum) and luminosities of the lobes at different observing frequencies. The lobes are approximated as a cylinder with length $D_{\ell}$ and diameter $D_{\ell}/AR$, where $AR$ is the axial ratio of the lobes.

In this paper, we follow the modelling procedure described in detail in \cite{machalski2007}. The assumed values of $D_{\ell}$ and $AR$ are given in Table~\ref{tab:GRG_properties}. We note that $D_{\ell}$ and $AR$ determine $V_{\ell}$, while the spectral index and luminosity provide the slope and power normalisation, respectively, of the spectrum at a particular frequency. To find the spectral indices required for the $k$-corrections, the simple analytical functions $y = a+bx+c \exp(\pm x)$, where $x = \log \nu {\rm [GHz]}$ and $y = \log S(\nu){\rm [mJy]}$, are fitted to the observed lobe flux densities, weighted by their respective errors. Accurate $k$-corrections are essential for calculating lobe luminosities at various observing frequencies. This fitting process yields model luminosities, which are subsequently converted into the model flux densities shown in column 6 of Table~\ref{tab:grq_flux_table}.

Following KDA, the modelling procedure expresses the initial ratio of the magnetic field energy density to relativistic particle energy density as $r=(1 + p)/4 = (1 + \alpha_{\rm inj})/2$. Under this formalism, the minimum predicted value for $r$ is 0.5, occurring when $\alpha_{\rm inj} = 0$. However, since the observed low-frequency spectral slopes of the GRQs indicate $0.5 \la \alpha_{\rm inj} \la 0.7$, the resulting fitted magnetic field strengths range from 0.75 to 0.85 of the equipartition value.

The results of the modelling are summarised in Table~\ref{tab:model_parameters}, which gives the four independent model parameters determined by the fit ($\alpha_{\rm inj}$, $t_{\ell}$, $Q_{\rm j}$ and $\rho_{0}$) as well as the following derivative parameters: 
\begin{itemize}
\item The sideways pressure in the lobes, $p_{\ell}$.
\item The ratio of the time-dependent expansion velocity of the jet head to the speed of light, $v_{\rm h}/c \equiv v_{\rm h}(t)/c$, where $v_{\rm h}(t) = \mathrm{d} D_{\ell}(t)/\mathrm{d} t$ and $D_{\ell}(t)$ is given by equation 1 of \cite{machalski2007}.
\item The ambient medium density at the head of the lobes, $\rho_{\rm X}$.
\item The magnetic field strength, $B_{\ell}$.
\item The energy of relativistic electrons within the lobes, $E_{\ell}$.
\item The rest-frame spectral index between 408 and 5000~MHz, $\alpha_{\rm em}$.
\end{itemize}

\begin{table*}
\caption{Model parameters. The columns give: (1) source name; (2) injection spectral index; (3) age of the lobes; (4) jet power; (5) medium density at the radio core; (6) pressure within the lobes; (7) ratio of the lobes' expansion speed to the speed of light; (8) ambient medium density at the head of the lobes; (9) magnetic field strength; (10) energy of relativistic electrons within the lobes; (11) two-point spectral index between 408 and 5000~MHz in the rest frame.}
\begin{tabular}{ccccccccccc}
\hline
Source name     & $\alpha_{\rm inj}$ & $t_{\ell}$ & $Q_{\rm j}$ & $\rho_{0}$ &
$p_{\ell}$ & $v_{\rm h}/c$ & $\rho_{\rm X}$ & $B_{\ell}$ & 
$E_{\ell}$ & $\alpha_{\rm em}$   \\
       &  & (Myr) &  ($10^{38}$\,W) & ($10^{-24}$\,kg\,m$^{-3}$) & ($10^{-13}$\,N\,m$^{-2}$) & &($10^{-26}$\,kg\,m$^{-3}$) & (nT) & ($10^{52}$\,J) \\
  (1)  &     (2)  &  (3)  &  (4)  &  (5)  &  (6)  &  (7)  & (8) &  (9) &  (10) & (11) \\
\hline
J1109+5104 & $0.62$ & $16.5$ & $3.7$ & $2.03$ & $1.32$   & $0.075$ & $0.69$ & $0.67$ & $6.2$ & $0.94$ \\
J1233+4902 & $0.55$ & $40$ & $1.21$ & $3.1$ & $0.78$   & $0.0290$ & $1.20$ & $0.51$ & $8.3$ & $1.00$ \\
J1238+5325 & $0.53$ & $91$ & $0.176$ & $1.26$ & $0.140$   & $0.0120$ & $0.50$ & $0.210$ & $3.8$ & $0.89$ \\
J1240+5334 & $0.64$ & $43$ & $0.62$ & $4.0$ & $0.71$   & $0.0230$ & $1.89$ & $0.49$ & $4.3$ & $0.86$ \\
J1326+5358 & $0.58$ & $213$ & $0.106$ & $96$ & $0.50$   & $0.0054$ & $36$ & $0.41$ & $2.90$ & $1.02$ \\
J1334+5501 & $0.58$ & $34$ & $10.3$ & $9.2$ & $4.9$   & $0.032$ & $3.9$ & $1.28$ & $72$ & $0.98$ \\
J1334+4813 & $0.60$ & $15.4$ & $17.4$ & $4.0$ & $5.5$   & $0.071$ & $1.61$ & $1.36$ & $39$ & $1.05$ \\
J1408+4738 & $0.55$ & $24.9$ & $3.0$ & $3.4$ & $1.98$   & $0.039$ & $1.62$ & $0.81$ & $13.0$ & $0.97$ \\
J1415+4909 & $0.54$ & $22.4$ & $2.31$ & $1.37$ & $0.80$   & $0.056$ & $0.45$ & $0.51$ & $7.3$ & $0.95$ \\
J1419+4837 & $0.68$ & $19.2$ & $2.07$ & $0.97$ & $0.91$   & $0.055$ & $0.42$ & $0.56$ & $6.5$ & $0.84$ \\
J1439+4550 & $0.51$ & $14.8$ & $5.8$ & $0.46$ & $1.03$   & $0.082$ & $0.161$ & $0.58$ & $16.1$ & $0.94$ \\
J1440+5026 & $0.62$ & $17.9$ & $7.3$ & $0.95$ & $0.92$   & $0.093$ & $0.206$ & $0.56$ & $17.7$ & $1.04$ \\
J1450+4549 & $0.61$ & $15.9$ & $11.5$ & $7.9$ & $5.3$   & $0.071$ & $3.1$ & $1.33$ & $18.9$ & $0.99$ \\
J1450+5300 & $0.63$ & $52$ & $6.0$ & $13.3$ & $1.44$   & $0.038$ & $2.25$ & $0.70$ & $32$ & $1.04$ \\
J1523+5203 & $0.72$ & $265$ & $0.44$ & $88$ & $0.94$   & $0.0047$ & $30$ & $0.57$ & $25.6$ & $1.19$ \\
\hline
\label{tab:model_parameters}
\end{tabular}
\end{table*}

In deriving $\rho_{\rm X}$, the density distribution of the unperturbed gas surrounding the RG is modelled with a $\beta$ profile
\begin{equation}
\rho(r)=\rho_{0}\left(\frac{r}{a_{0}}\right)^{-\beta} \, ,
\label{eq:rho(r)}
\end{equation}
where $r > a_{0}$ is the radial distance from the centre of the host galaxy, $a_{{0}} = 10$~kpc, $\beta = 3/2$ and $\rho_{\rm X}=\rho(D_{\ell}/2)$. The choice of these parameter values is justified in \cite{machalski2007}; the same parameter values were adopted by \cite{machalski2021} in their analysis of the Atlas sample. The rest-frame spectral index between 408 and 5000~MHz is calculated as outlined in \cite{machalski2021}. The equations used to derive the other derivative parameters are given in \cite{machalski2007} and \cite{machalski2011a}.

To aid with the interpretation of the results, we also assess the impact of the two largest assumptions in the model parameters. Firstly, it is important to emphasise that $\rho_{0}$ and $\beta$ are not independent parameters; the model fit to the observables provides the best-fit value for the term $\rho_{0}\,a_{0}^{\beta}$, not for $\rho_{0}$ alone. To evaluate the impact of this, in Appendix~\ref{Refitting}, we investigate the sensitivity of $t_{\ell}$, $Q_{\rm j}$, $\rho_{0}$, $\rho_{\rm X}$, $\alpha_{\rm em}$, $v_{\rm h}/c$ and $B_{\ell}$ to the assumed $\beta$ for our GRQ sample and find that only $\rho_{0}$ is strongly dependent on $\beta$: a variation in $\beta$ over a wide range (0.5 to 1.9) results in an increase in $\rho_{0}$ by a factor of $\approx 50-200$. We note that $\rho_{\rm X}$ is only weakly dependent on $\beta$ as a consequence of the relationship between $\rho_{\rm X}$ and $\rho_{0}$. Therefore, all the aforementioned fitted model parameters, except $\rho_{0}$, can be reliably used in this statistical study.

Secondly, as discussed in Section~\ref{Radio morphology}, most of the GRQs likely lie close to the plane of the sky based on their radio morphologies in FIRST and VLASS, and $i$ is assumed to be either 90 or 70~deg. For completeness, we explore the effect of a decrease in $i$ from 90 to 45~deg on the results from the dynamical modelling. In the frame of the DYNAGE algorithm, this change in $i$ results in an inevitable increase in the observables $D_{\ell}$ and $AR$, and an increase in $t_{\ell}$, $Q_{\rm j}$, and $\rho_{0}$ by a factor of $\approx 1.05{-}1.1$, $\approx 1.35{-}1.5$ and $\approx 1.1{-}1.2$, respectively. 


\section{Statistical implications}

The dynamical evolution models of statistically large samples of extended radio sources can be used to analyse several potential correlations between the model parameters. This approach has been possible since the publication of DYNAGE models for the Atlas \citep{machalski2021}. In particular, \cite{wojtowicz2021} showed that detailed modelling of the non-thermal emission from the radio lobes allows some of those parameters to be constrained, and noticed significant and non-obvious correlations between $\log \rho_{\rm X}$ and $\alpha_{\rm em}$ combined with $\log D_{\ell}$, and $\log \rho_{0}$ and $\alpha_{\rm em}$. They derived the corresponding correlation parameters and quantified the intrinsic scatter using a Bayesian analysis.

In this Section, we investigate whether or not the above correlations for the Atlas RGs and RQs are significantly different, and attempt to improve the correlations by adding the GRQs studied in this paper to the Atlas RQs.

\begin{figure*}
\begin{center}
\includegraphics[scale=0.80, angle=0, trim=0cm 0cm 0cm 0cm]{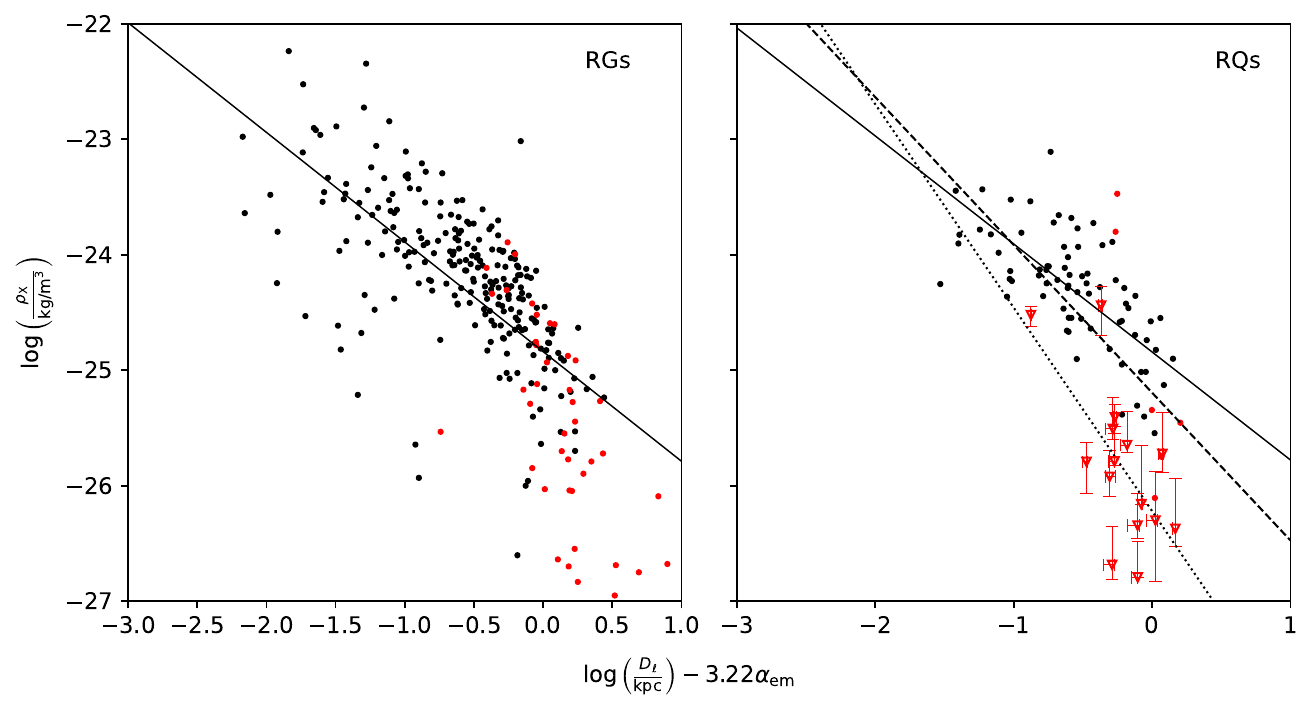}
\caption{Correlation between the ambient medium density at the head of the lobes and the linear extent of the lobes combined with the rest-frame two-point spectral index between 408 and 5000~MHz. Left: Non-giant RGs (black dots) and GRGs (red dots) from the Atlas. The solid line is the least-squares linear regression line. Right: Non-giant RQs (black dots) and GRQs (red dots) from the Atlas, and GRQs modelled in this paper (red triangles). The error bars for our GRQs correspond to the parameter ranges yielded by varying $\beta$ from 0.5 to 1.9. The solid line is the regression line for the Atlas RQs alone, the dashed line is the regression line for both the Atlas RQs and our GRQs, and the dotted line is the regression line for our GRQs alone. The correlation for each source subset is quantified in columns 3--5 of Table~\ref{tab:correlation}.}
\label{fig:correlation_lobes}
\end{center}
\end{figure*}

\begin{figure*}
\begin{center}
\includegraphics[scale=0.80, angle=0, trim=0cm 0cm 0cm 0cm]{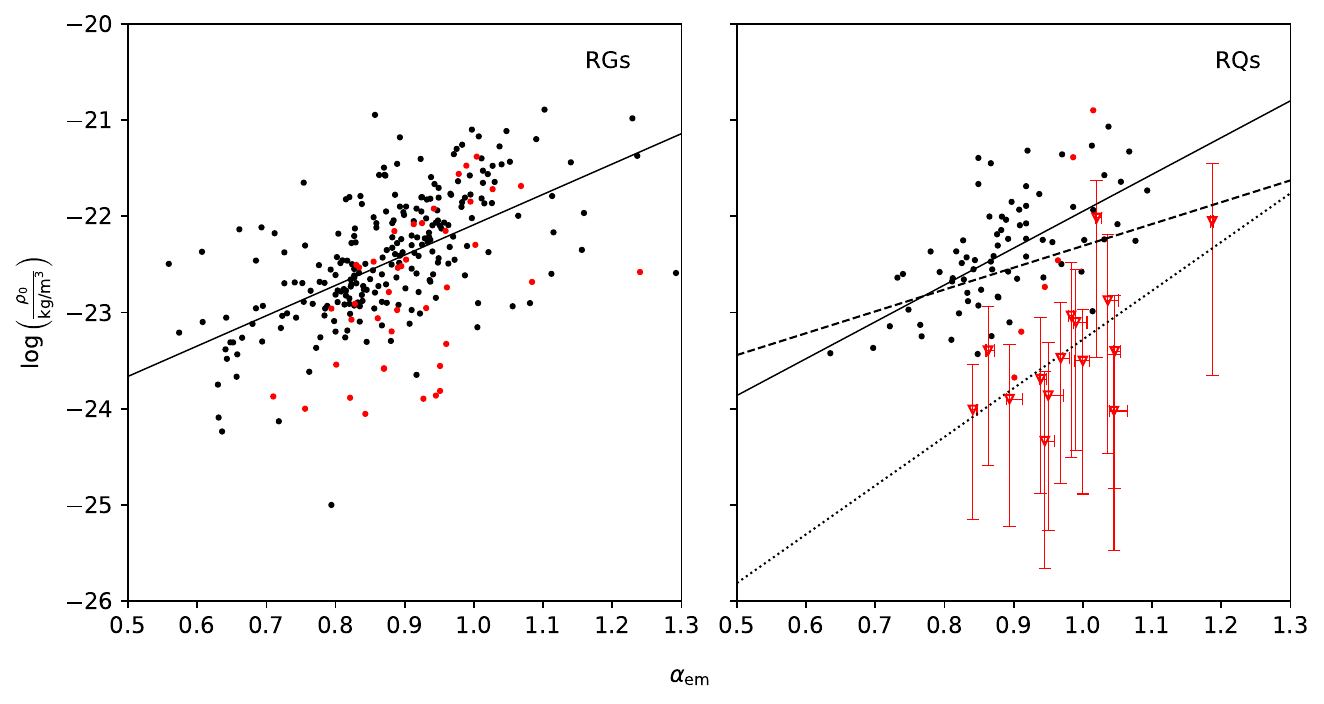}
\caption{Correlation between the medium density at the core radius and the rest-frame spectral index between 408 and 5000~MHz. Left: Non-giant RGs (black dots) and GRGs (red dots) from the Atlas. The solid line is the least-squares linear regression line. Right: Non-giant RQs (black dots) and GRQs (red dots) from the Atlas, and GRQs modelled in this paper (red triangles). The error bars for our GRQs correspond to the parameter ranges yielded by varying $\beta$ from 0.5 to 1.9. The solid line is the regression line for the Atlas RQs alone, the dashed line is the regression line for both the Atlas RQs and our GRQs, and the dotted line is the regression line for our GRQs alone. The correlation for each source subset is quantified in columns 6--8 of Table~\ref{tab:correlation}.}
\label{fig:correlation_core}
\end{center}
\end{figure*}

Following the approach by \cite{wojtowicz2021}, we consider the following two correlations: $\log \rho_{\rm X}$ versus $\log D_{\ell} - 3.22 \alpha_{\rm em}$ (see their fig.~2) and $\log \rho_{0}$ versus $\alpha_{\rm em}$ (see their fig.~6). These two correlations are presented in Figs~\ref{fig:correlation_lobes} and \ref{fig:correlation_core}; in the left panels, they are shown for the Atlas RGs, and in the right panels for the Atlas RQs supplemented with our GRQs. In both figures, the solid lines indicate the least-squares linear regression lines for the Atlas data points, the dashed lines the regression lines for the Atlas RQs combined with our GRQs and the dotted lines the regression lines for our GRQs alone. The error bars for our GRQs reflect the parameter ranges obtained by varying $\beta$ between 0.5 and 1.9; the uncertainty on $\rho_{0}$ is particularly large due to the factors detailed in Section~\ref{Dynamical modelling}.

The correlations for the different source subsets are quantified in Table~\ref{tab:correlation}. For both correlations, there is no significant difference between the regression lines for the Atlas RGs and RQs. This is consistent with AGN unification models which predict that RGs and RQs can be unified as a single class based on the fact that they have the same intrinsic properties.

\begin{table*}
\centering
\caption{Results of the two correlation analyses for the different source subsets. $N$ is the number of sources. The first correlation is between $x \equiv \log\,D_{\ell}-3.22\alpha_{\rm em}$ and $y \equiv \log\rho_{\rm X}$ and the second correlation is between $x \equiv \alpha_{\rm em}$ and $y \equiv \log\rho_{0}$. The linear regression line is given by $ax+b$, $R$ is the Pearson correlation coefficient and $P$ is the probability of the null hypothesis that the slope is zero.}
\label{tab:correlation}
\begin{tabular}{@{} c c | c c c c | c c c c}
\hline
& & \multicolumn{4}{c|}{Correlation 1} & \multicolumn{4}{c}{Correlation 2} \\
\cmidrule(lr){3-6}\cmidrule(lr){7-10}
Population & $N$ & $a$ & $b$ & $R$ & $P$ & $a$ & $b$ & $R$ & $P$ \\
\hline
Atlas galaxies & $ 281 $ & $ -0.95 \pm 0.06 $ & $ 0.16 \pm 0.05 $ & $ -0.68 $ & $3.23 \times 10^{-40}$ & $ 3.15 \pm 0.27 $ & $ -2.24 \pm 0.24 $ & $ 0.57 $ & $7.60 \times 10^{-26}$ \\
Atlas quasars & $ 80 $ & $ -0.94 \pm 0.12 $ & $ 0.16 \pm 0.08 $ & $ -0.67 $ & $1.44 \times 10^{-11}$ & $ 3.83 \pm 0.60 $ & $ -2.78 \pm 0.54 $ & $ 0.59 $ & $1.04 \times 10^{-8}$ \\
GRQs & $ 15 $ & $ -1.76 \pm 0.58 $ & $ -1.22 \pm 0.19 $ & $ -0.65 $ & $9.08 \times 10^{-3}$ & $ 5.06 \pm 1.71 $ & $ -5.34 \pm 1.68 $ & $ 0.64 $ & $1.09 \times 10^{-2}$ \\
Atlas quasars + GRQs & $ 95 $ & $ -1.28 \pm 0.16 $ & $ -0.19 \pm 0.10 $ & $ -0.65 $ & $1.59 \times 10^{-12}$ & $ 2.27 \pm 0.73 $ & $ -1.58 \pm 0.67 $ & $ 0.31 $ & $2.63 \times 10^{-3}$ \\
\hline
\end{tabular}
\end{table*}

For both correlations, the regression lines for our GRQs lie significantly below those for the Atlas RQs. When combining the two samples, the slope of the correlation between $\log \rho_{\rm X}$ and $\log D_{\ell} - 3.22 \alpha_{\rm em}$ steepens by $\approx 0.3$ with almost no change in the correlation coefficient. The slope of the correlation between $\log \rho_{0}$ and $\alpha_{\rm em}$ flattens by $\approx 1.6$ and the correlation coefficient drops by $\approx 0.3$.

The Atlas includes six GRQs and 41 GRGs with $D_{\ell} \geq 0.7$~Mpc; these sources are represented as red dots in Figs~\ref{fig:correlation_lobes} and \ref{fig:correlation_core}. A similar trend is observed for the GRGs in Atlas: for both correlations, about 70 per cent of the GRGs lie below the regression lines for the RGs. There are too few GRQs in Atlas to be able to confirm the trend.


\section{Discussion}

\subsection{Exploring the relationship between ambient medium density and linear size}

The correlations between $\log \rho_{\rm X}$ and $\log D_{\ell} - 3.22 \alpha_{\rm em}$, and $\log \rho_{0}$ and $\alpha_{\rm em}$, determined for our sample of 15 GRQs, do not closely follow the correlations arising from the Atlas RQs. Both densities, $\rho_{X}$ and $\rho_{0}$, tend to be significantly lower than those of the Atlas RQs. These results are consistent with other studies of samples of GRSs \citep[e.g.][]{mack1998,schoenmakers2000,stuardi2020} that have shown that their growth is favoured in under-dense environments (particle densities in front of the lobes of less than a few times $10^{-5}~\mathrm{cm}^{-3}$ and magnetic fields $B_{\ell} \lesssim \,1$~nT)\footnote{A particle density of $10^{-5}~\mathrm{cm}^{-3}$ corresponds to a mass density of $2.32 \cdot 10^{-26}~\mathrm{kg}~\mathrm{m}^{-3}$ assuming a mean particle mass of 1.4 amu.}.

We caution that this finding depends on the assumption that $\beta = 1.5$ for both our GRQs and the Atlas RQs. \cite{vikhlinin2006} demonstrated that gas density profiles in relaxed galaxy clusters undergo continuous steepening with radius; consequently, our GRQs likely reside in steeper sections of the ambient medium than the Atlas RQs. As shown in Appendix~\ref{beta_from_xray}, adopting the mean cluster profile from \citeauthor{vikhlinin2006} yields $\beta \approx 0.8-1.1$ for the non-giant RQs in Atlas and $\approx 1.1-1.2$ for our GRQs -- values significantly flatter than the $\beta = 1.5$ assumed in our dynamical modelling. We show in Appendix~\ref{Refitting} that while $\rho_{0}$ is strongly dependent on the assumed $\beta$, an overestimation of the profile steepness leads to an overestimate of $\rho_{0}$. While incorporating more realistic, non-power-law density profiles would permit a more robust comparison of central core densities, such an analysis is beyond the scope of the current work.

\begin{figure*}
\begin{center}
\includegraphics[scale=0.56, angle=0, trim=0cm 0cm 0cm 0cm]{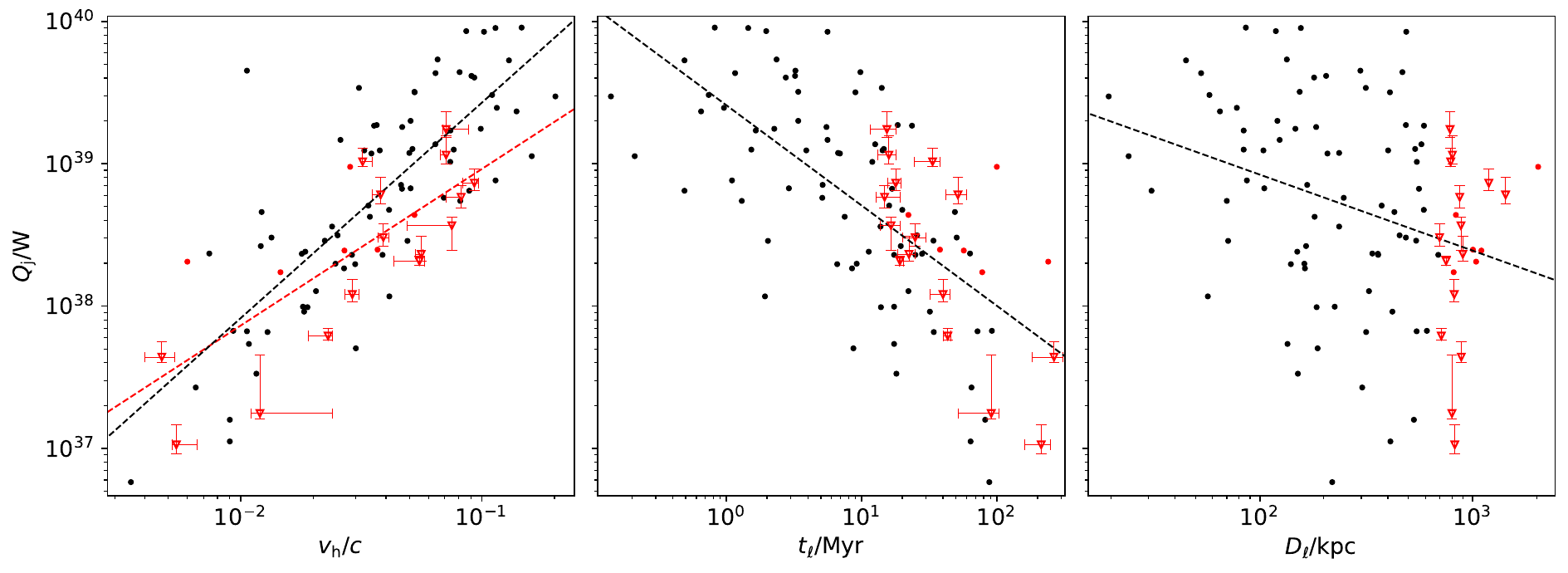}
\caption{Jet power versus jet advance velocity (left), age of the lobes (centre) and deprojected linear size (right) for our GRQs (red triangles), the Atlas GRQs (red dots) and non-giant RQs (black dots). The error bars for the GRQs modelled in this paper correspond to the parameter ranges yielded by varying $\beta$ from 0.5 to 1.9. In the left panel, the black dashed line is the regression line for the non-giant RQs and the red dashed line the regression line for the GRQs. In the other panels, the black dashed lines are the regression lines for all RQs.}
\label{fig:jet_power}
\end{center}
\end{figure*}

Another hypothesis that has been proposed to explain the large sizes of GRSs is that they possess more powerful radio jets than normal RGs, allowing them to reach Mpc scales \citep{wiita1989}. The left panel of Fig.~\ref{fig:jet_power} shows $Q_{\rm j}$ as a function of $v_{\rm h}/c$ for all RQs. The mean jet power of the GRQs ($10^{ 38.39  \pm  0.12}~{\rm W}$) is slightly lower than that of the smaller-sized RQs ($10^{ 38.77 \pm 0.08 }~{\rm W}$). A high jet power therefore does not appear to be necessary to explain the large sizes of the GRQs. A similar result was obtained by \cite{gurkan2022} who examined the radio power-linear size ($P{-}D$) distribution of resolved AGN detected with the Australian SKA Pathfinder \citep[ASKAP;][]{hotan2021} in the Galaxy and Mass Assembly \citep[GAMA;][]{driver2016} 23 field, including 63 GRSs. The GRSs were found to have similar jet powers to smaller-sized jetted AGN.

\cite{oei2024} measured multi-Mpc-scale Cosmic Web densities around the hosts of $\sim 100$ luminous ($P_{150~{\rm MHz}} > 10^{24}~{\rm W}~{\rm Hz}^{-1}$) GRSs, mostly discovered in LoTSS DR2, and $\sim 1000$ LoTSS DR1 smaller-sized radio sources with typical 150~MHz radio luminosities in the range $10^{22}{-}10^{24}~{\rm W}~{\rm Hz}^{-1}$. The GRSs were found to inhabit denser regions of the Cosmic Web than the control radio source sample. Since radio luminosity is a proxy for jet power, their interpretation is that these sources are able to overcome the IGM's resistance to their Mpc-scale growth thanks to their high jet powers. As previously noted, the central core densities of the GRQs in this study cannot be robustly compared with those of the non-giant RQs without further refinements to the dynamical modelling. However, given that our GRQ sample exhibits slightly lower average jet powers, it is plausible that they evolve in environments with lower medium densities -- a result that contrasts with \citeauthor{oei2024}, where higher jet powers were associated with denser regions.

\subsection{Correlations between jet power and jet velocity, lobe age and linear size}

We find a strong correlation between $Q_{\rm j}$ and $v_{\rm h}/c$ for both the GRQs and smaller-sized RQs, as shown in the left panel of Fig.~\ref{fig:jet_power}; the correlations are quantified in Table~\ref{tab:jet_power_correlation}. This is consistent with previous observational evidence for a correlation between the radio power and expansion speed of a radio source's lobes \citep[e.g.][]{alexander1987,liu1992}. The result is entirely expected and aligns with classical models of radio galaxy growth, wherein the expansion of the radio lobe is governed by the balancing pressure between the momentum flux of the jet and the ram pressure of the unperturbed external medium. The high Pearson correlation coefficient of $R = 0.77$ for all RQs indicates that the thrust generated by the central engine's jet power is the primary driver regulating the kinetic speed at which the radio structure expands into the IGM, overriding minor environmental fluctuations. We note that the ages of our GRQs ($14.8{-}265$~Myr) always correspond to `realistic' values of their jet advance velocities ($0.093{-}0.0047c$). Indeed, jet advance velocities as high as $\approx 0.3c$ have been observed for young radio sources using the VLBI technique \citep[see e.g.][]{polatidis2003}, and \cite{scheuer1995} gives an upper limit of $0.1 c$ for the velocities of the jets in `regular' FRII RGs.

\begin{table*}
\centering
\caption{Correlation between $Q_{\rm j}$ and $v_{\rm h}/c$ for the GRQs, non-giant RQs and all RQs. $N$ is the number of sources. The linear regression line is given by $ax+b$, $R$ is the Pearson correlation coefficient and $P$ is the probability of the null hypothesis that the slope is zero.}
\label{tab:jet_power_correlation}
\begin{tabular}{@{} c c | c c c c | c c c c}
\hline
& & \multicolumn{4}{c}{Correlation between $Q_{\rm j}$ and $v_{\rm h}/c$} \\
\cmidrule(lr){3-6}
Population & $N$ & $a$ & $b$ & $R$ & $P$ \\
\hline
GRQs & $ 21 $ & $ 1.10 \pm 0.23 $ & $ 2.06 \pm 0.36 $ & $ 0.74 $ & $1.20 \times 10^{-4}$ \\
Normal-sized RQs & $ 74 $ & $ 1.51 \pm 0.14 $ & $ 2.94 \pm 0.21 $ & $ 0.79 $ & $6.88 \times 10^{-17}$ \\
All RQs & $ 95 $ & $ 1.45 \pm 0.12 $ & $ 2.79 \pm 0.18 $ & $ 0.77 $ & $3.20 \times 10^{-20}$ \\
\hline
\end{tabular}
\end{table*}

The middle and right panels of Fig.~\ref{fig:jet_power} show $Q_{\rm j}$ as a function of $t_{\ell}$ and $D_{\ell}$, respectively, for all RQs; the correlations are quantified in Table~\ref{tab:correlation_spearman}. The jet power is strongly anti-correlated with $t_{\ell}$ ($R = -0.67$), which we attribute to the well-documented observational phenomenon known as the youth-redshift degeneracy or youth-luminosity selection effect \citep{blundell1999b}. As radio lobes age and expand to larger volumes, they suffer significant adiabatic expansion losses and synchrotron cooling, causing their radio luminosity to drop rapidly. Consequently, highly powerful jets evolve rapidly and are detected preferentially when they are dynamically young.

While $Q_{\rm j}$ is strongly tied to the age of the system, it displays a weak anti-correlation with $D_{\ell}$ ($R = -0.32$). Dynamically, $D_{\ell}$ is the time-integral of $v_{\rm h}/c$. Because high jet power drives a faster expansion speed but restricts our detection window to younger systems due to the aforementioned youth bias, these two physical mechanisms act in direct competition, masking the true physical relationship between $Q_{\rm j}$ and $D_{\ell}$.

To disentangle these competing effects, we perform a Spearman partial rank correlation analysis \citep{macklin1982} between $Q_{\rm j}$ and $D_{\ell}$ while holding $t_{\ell}$ constant, the results of which are shown in columns 6--8 of Table~\ref{tab:correlation_spearman}. This analysis reveals a significant, positive partial rank correlation coefficient ($r = 0.51$). By controlling for $t_{\ell}$, we eliminate the youth bias, demonstrating that for sources of an identical evolutionary age, higher jet powers do indeed yield larger physical extents.

\begin{table*}
\centering
\caption{Correlations between $Q_{\rm j}$ and $t_{\ell}$, $Q_{\rm j}$ and $D_{\ell}$, and $t_{\ell}$ and $D_{\ell}$, for all RQs. Columns 2--5 show the results of a linear least-squares regression analysis. The linear regression line is given by $ax+b$ and $R$ is the Pearson correlation coefficient. Columns 6--8 show the results of a Spearman partial rank correlation analysis: $r_{AX}$ is the Spearman correlation coefficient between $A$ and $X$, $r_{AX,Y}$ is the Spearman partial rank correlation coefficient between $A$ and $X$ when $Y$ is held constant, and $D_{AX,Y}$ is the associated significance level.}
\label{tab:correlation_spearman}
\begin{tabular}{@{} c | c c c c | c c c}
\hline
& \multicolumn{4}{c|}{Linear least-squares regression analysis} & \multicolumn{3}{c}{Spearman partial rank correlation analysis} \\
\cmidrule(lr){2-5}\cmidrule(lr){6-8}
Correlation & $a$ & $b$ & $R$ & $P$ & $r_{AX}$ & $r_{AX,Y}$ & $D_{AX,Y}$ \\
\hline
 $Q_{\mathrm{j}} t_{\ell}$ & $ -0.70 \pm 0.08 $ & $ 1.41 \pm 0.10 $ & $ -0.67 $ & $1.49 \times 10^{-13}$ & $ -0.69 $ & $ -0.75 $ & $ -9.21 $ \\ 
 $Q_{\mathrm{j}} D_{\ell}$ & $ -0.53 \pm 0.17 $ & $ 1.99 \pm 0.41 $ & $ -0.32 $ & $1.73 \times 10^{-3}$ & $ -0.34 $ & $ 0.51 $ & $ 5.38 $ \\ 
 $t_{\ell} D_{\ell}$ & $ 1.36 \pm 0.09 $ & $ -2.30 \pm 0.21 $ & $ 0.85 $ & $3.45 \times 10^{-28}$ & $ 0.81 $ & $ 0.84 $ & $ 11.80 $ \\
\hline
\end{tabular}
\end{table*}

We also perform a partial rank correlation analysis between $Q_{\rm j}$ and $t_{\ell}$ while holding $D_{\ell}$ constant and find that the anti-correlation becomes even more pronounced, tightening to $r = -0.75$. This result is expected when considering a fixed $D_{\ell}$: for a population of sources that have expanded to the exact same physical extent, those driven by a higher jet power will tend to propagate at higher advance velocities -- and thus require a shorter duration to reach that size -- subject to variations in their local ambient environments. By controlling for $D_{\ell}$, we eliminate the statistical variance introduced by the wide range of linear sizes.


\section{Conclusions}

We present dynamical evolution models of 15 GRQs with linear sizes $\geq 0.7$~Mpc in the HETDEX Spring Field. Their observational data, including the flux densities and spectral indices measured at a number of observing frequencies between 54{-}4850 MHz, are compiled from existing radio images and catalogues including LoLSS DR1, LoTSS DR2, NVSS, FIRST and VLASS. For four of the sources, the radio spectra are constrained up to 15.5~GHz using follow-up observations with the AMI-LA. Most of the GRQs have quite symmetrical radio structures in the FIRST and VLASS images above 1~GHz, indicating that they lie close to the plane of the sky. Using the DYNAGE algorithm, we derive estimates of their physical parameters, including the age of the lobes, jet power and ambient medium density.

We investigate previously studied correlations between the ambient medium density and observed radio spectrum for a compiled sample of FRII-type radio sources with a large range of deprojected linear sizes, $D_{\ell}$ (9.8~kpc to 2.58~Mpc). Following the approach by \cite{wojtowicz2021}, we consider the following two correlations: $\log \rho_{\rm X}$ versus $\log D_{\ell} - 3.22 \alpha_{\rm em}$ and $\log \rho_{0}$ versus $\alpha_{\rm em}$, where $\rho_{\rm X}$ is the ambient medium density at the head of the lobes, $\rho_{0}$ the central density at the core radius and $\alpha_{\rm em}$ the rest-frame spectral index between 408 and 5000~MHz. For both correlations, we find that the regression lines for the RGs and RQs in the Atlas are almost identical, in line with AGN unification models.

Our sample of GRQs forms an outlier population with respect to both correlations: they lie in low-density environments both at the central core radius ($\rho_{0} \lesssim 10^{-23}~\mathrm{kg}~\mathrm{m}^{-3}$) and in front of the lobes ($\rho_{\rm X} \lesssim 10^{-25}~\mathrm{kg}~\mathrm{m}^{-3}$). They are not exceptionally old ($t_{\ell} \lesssim 100$~Myr) compared with the smaller-sized RQs in Atlas and have similar jet powers ($Q_{\rm j}$ in the range $\sim 10^{37}{-}10^{39}$~W). Our interpretation is that the low medium density in which they evolve enable them to grow faster and reach Mpc scales within a few tens of Myr.

There is a strong correlation ($R = 0.77$, where $R$ is the Pearson correlation coefficient) between $Q_{\rm j}$ and the ratio of the expansion velocity of the jet head to the speed of light, $v_{\rm h}/c$, for the Atlas RQs combined with our GRQs. The jet power is anti-correlated with both $t_{\ell}$ ($R=-0.67$) and $D_{\ell}$ ($R=-0.32$), but $t_{\ell}$ is a much stronger indicator of $Q_{\rm j}$ than $D_{\ell}$, even though the age fit parameter is less certain than the observed angular size, hence the source's linear extent. We attribute the strong anti-correlation between $Q_{\rm j}$ and $t_{\ell}$ to the youth-luminosity selection effect, whereby flux-limited surveys preferentially detect younger, highly compact radio sources during their peak luminosity phase before severe cooling and expansion losses cause them to fade. Using a Spearman partial rank correlation analysis, we isolate a statistically significant anti-correlation between $Q_{\rm j}$ and $t_{\ell}$ at fixed $D_{\ell}$ ($r=-0.75$, where $r$ is the partial rank correlation coefficient). We also find a significant correlation between $Q_{\rm j}$ and $D_{\ell}$ at fixed $t_{\ell}$ ($r=0.51$). These results demonstrate a fundamental relation between $Q_{\rm j}$, $t_{\ell}$, and $D_{\ell}$, despite the strong underlying correlation between age and size.

Finally, we caution that the derived values of $\rho_{0}$ for both our GRQs and the Atlas RQs are highly sensitive to the assumed value of $\beta$, the slope of the power-law density profile for the undisturbed gas surrounding the source. A $\beta$ value of 1.5 was assumed in the dynamical modelling for both samples, while the density profile of clusters is known to steepen with radius. \cite{oei2023} have compiled a catalogue of 2060 GRS candidates discovered in LoTSS DR2 over an area of 5635\sqdeg, covering the projected length range $0.7{-}5.07$~Mpc. In future, our study could be extended to these sources, with improved estimates of $\rho_{0}$ using constraints on $\beta$ from X-ray measurements of the cluster gas.


\section*{Acknowledgements}

We thank the anonymous referees for helpful comments, which have substantially improved this paper. We thank Dave Titterington for his help scheduling the AMI observations and analysing the AMI data. We thank the staff of the Mullard Radio Astronomy Observatory, University of Cambridge, for their support in the maintenance, and operation of AMI. We acknowledge support from the European Research Council under grant ERC-2012-StG-307215 LODESTONE. This research has made use of the CIRADA cutout service at URL cutouts.cirada.ca, operated by the Canadian Initiative for Radio Astronomy Data Analysis (CIRADA). CIRADA is funded by a grant from the Canada Foundation for Innovation 2017 Innovation Fund (Project 35999), as well as by the Provinces of Ontario, British Columbia, Alberta, Manitoba and Quebec, in collaboration with the National Research Council of Canada, the US National Radio Astronomy Observatory and Australia’s Commonwealth Scientific and Industrial Research Organisation. This research has made use of data from the OVRO 40-m monitoring programme \citep{richards2011}, supported by private funding from the California Institute of Technology and the Max Planck Institute for Radio Astronomy, and by NASA grants NNX08AW31G, NNX11A043G, and NNX14AQ89G and NSF grants AST-0808050 and AST-1109911.


\section*{Data Availability}
The data underlying this article are available in the article.


\bibliographystyle{mnras}
\bibliography{myreferences} 

\begin{thebibliography}{}
\makeatletter
\relax
\def\mn@urlcharsother{\let\do\@makeother \do\$\do\&\do\#\do\^\do\_\do\%\do\~}
\def\mn@doi{\begingroup\mn@urlcharsother \@ifnextchar [ {\mn@doi@}
  {\mn@doi@[]}}
\def\mn@doi@[#1]#2{\def\@tempa{#1}\ifx\@tempa\@empty \href
  {http://dx.doi.org/#2} {doi:#2}\else \href {http://dx.doi.org/#2} {#1}\fi
  \endgroup}
\def\mn@eprint#1#2{\mn@eprint@#1:#2::\@nil}
\def\mn@eprint@arXiv#1{\href {http://arxiv.org/abs/#1} {{\tt arXiv:#1}}}
\def\mn@eprint@dblp#1{\href {http://dblp.uni-trier.de/rec/bibtex/#1.xml}
  {dblp:#1}}
\def\mn@eprint@#1:#2:#3:#4\@nil{\def\@tempa {#1}\def\@tempb {#2}\def\@tempc
  {#3}\ifx \@tempc \@empty \let \@tempc \@tempb \let \@tempb \@tempa \fi \ifx
  \@tempb \@empty \def\@tempb {arXiv}\fi \@ifundefined
  {mn@eprint@\@tempb}{\@tempb:\@tempc}{\expandafter \expandafter \csname
  mn@eprint@\@tempb\endcsname \expandafter{\@tempc}}}

\bibitem[\protect\citeauthoryear{{Ahumada} et~al.,}{{Ahumada}
  et~al.}{2020}]{ahumada2020}
{Ahumada} R.,  et~al., 2020, \mn@doi [\apjs] {10.3847/1538-4365/ab929e}, \href
  {https://ui.adsabs.harvard.edu/abs/2020ApJS..249....3A} {249, 3}

\bibitem[\protect\citeauthoryear{{Alexander} \& {Leahy}}{{Alexander} \&
  {Leahy}}{1987}]{alexander1987}
{Alexander} P.,  {Leahy} J.~P.,  1987, \mn@doi [\mnras]
  {10.1093/mnras/225.1.1}, \href
  {https://ui.adsabs.harvard.edu/abs/1987MNRAS.225....1A} {225, 1}

\bibitem[\protect\citeauthoryear{{Almeida} et~al.,}{{Almeida}
  et~al.}{2023}]{almeida2023}
{Almeida} A.,  et~al., 2023, \mn@doi [\apjs] {10.3847/1538-4365/acda98}, \href
  {https://ui.adsabs.harvard.edu/abs/2023ApJS..267...44A} {267, 44}

\bibitem[\protect\citeauthoryear{{Amirkhanyan}}{{Amirkhanyan}}{2016}]{amirkhanyan2016}
{Amirkhanyan} V.~R.,  2016, \mn@doi [Astrophysical Bulletin]
  {10.1134/S1990341316040027}, \href
  {https://ui.adsabs.harvard.edu/abs/2016AstBu..71..384A} {71, 384}

\bibitem[\protect\citeauthoryear{{Antonucci}}{{Antonucci}}{1993}]{antonucci1993}
{Antonucci} R.,  1993, \mn@doi [\araa] {10.1146/annurev.aa.31.090193.002353},
  \href {https://ui.adsabs.harvard.edu/abs/1993ARA&A..31..473A} {31, 473}

\bibitem[\protect\citeauthoryear{{Bardeen} \& {Petterson}}{{Bardeen} \&
  {Petterson}}{1975}]{bardeen1975}
{Bardeen} J.~M.,  {Petterson} J.~A.,  1975, \mn@doi [\apjl] {10.1086/181711},
  \href {https://ui.adsabs.harvard.edu/abs/1975ApJ...195L..65B} {195, L65}

\bibitem[\protect\citeauthoryear{{Becker}, {White}  \& {Edwards}}{{Becker}
  et~al.}{1991}]{becker1991}
{Becker} R.~H.,  {White} R.~L.,   {Edwards} A.~L.,  1991, \mn@doi [\apjs]
  {10.1086/191529}, \href
  {https://ui.adsabs.harvard.edu/abs/1991ApJS...75....1B} {75, 1}

\bibitem[\protect\citeauthoryear{{Becker}, {White}  \& {Helfand}}{{Becker}
  et~al.}{1995}]{becker1995}
{Becker} R.~H.,  {White} R.~L.,   {Helfand} D.~J.,  1995, \mn@doi [\apj]
  {10.1086/176166}, \href
  {https://ui.adsabs.harvard.edu/abs/1995ApJ...450..559B} {450, 559}

\bibitem[\protect\citeauthoryear{{Begelman} \& {Cioffi}}{{Begelman} \&
  {Cioffi}}{1989}]{begelman1989}
{Begelman} M.~C.,  {Cioffi} D.~F.,  1989, \mn@doi [\apjl] {10.1086/185542},
  \href {https://ui.adsabs.harvard.edu/abs/1989ApJ...345L..21B} {345, L21}

\bibitem[\protect\citeauthoryear{{Best} \& {Heckman}}{{Best} \&
  {Heckman}}{2012}]{best2012}
{Best} P.~N.,  {Heckman} T.~M.,  2012, \mn@doi [\mnras]
  {10.1111/j.1365-2966.2012.20414.x}, \href
  {https://ui.adsabs.harvard.edu/abs/2012MNRAS.421.1569B} {421, 1569}

\bibitem[\protect\citeauthoryear{{Blundell} \& {Rawlings}}{{Blundell} \&
  {Rawlings}}{1999}]{blundell1999b}
{Blundell} K.~M.,  {Rawlings} S.,  1999, \mn@doi [\nat] {10.1038/20612}, \href
  {https://ui.adsabs.harvard.edu/abs/1999Natur.399..330B} {399, 330}

\bibitem[\protect\citeauthoryear{{Blundell}, {Rawlings}  \&
  {Willott}}{{Blundell} et~al.}{1999}]{blundell1999}
{Blundell} K.~M.,  {Rawlings} S.,   {Willott} C.~J.,  1999, \mn@doi [\aj]
  {10.1086/300721}, \href
  {https://ui.adsabs.harvard.edu/abs/1999AJ....117..677B} {117, 677}

\bibitem[\protect\citeauthoryear{{Cohen}, {Lane}, {Cotton}, {Kassim}, {Lazio},
  {Perley}, {Condon}  \& {Erickson}}{{Cohen} et~al.}{2007}]{cohen2007}
{Cohen} A.~S.,  {Lane} W.~M.,  {Cotton} W.~D.,  {Kassim} N.~E.,  {Lazio}
  T.~J.~W.,  {Perley} R.~A.,  {Condon} J.~J.,   {Erickson} W.~C.,  2007,
  \mn@doi [\aj] {10.1086/520719}, \href
  {https://ui.adsabs.harvard.edu/abs/2007AJ....134.1245C} {134, 1245}

\bibitem[\protect\citeauthoryear{{Condon}, {Cotton}, {Greisen}, {Yin},
  {Perley}, {Taylor}  \& {Broderick}}{{Condon} et~al.}{1998}]{condon1998}
{Condon} J.~J.,  {Cotton} W.~D.,  {Greisen} E.~W.,  {Yin} Q.~F.,  {Perley}
  R.~A.,  {Taylor} G.~B.,   {Broderick} J.~J.,  1998, \mn@doi [\aj]
  {10.1086/300337}, \href
  {https://ui.adsabs.harvard.edu/abs/1998AJ....115.1693C} {115, 1693}

\bibitem[\protect\citeauthoryear{{Dabhade} et~al.,}{{Dabhade}
  et~al.}{2020}]{dabhade2020}
{Dabhade} P.,  et~al., 2020, \mn@doi [\aap] {10.1051/0004-6361/201935589},
  \href {https://ui.adsabs.harvard.edu/abs/2020A&A...635A...5D} {635, A5}

\bibitem[\protect\citeauthoryear{{Dav{\'e}} et~al.,}{{Dav{\'e}}
  et~al.}{2001}]{dave2001}
{Dav{\'e}} R.,  et~al., 2001, \mn@doi [\apj] {10.1086/320548}, \href
  {https://ui.adsabs.harvard.edu/abs/2001ApJ...552..473D} {552, 473}

\bibitem[\protect\citeauthoryear{{Driver} et~al.,}{{Driver}
  et~al.}{2016}]{driver2016}
{Driver} S.~P.,  et~al., 2016, \mn@doi [\mnras] {10.1093/mnras/stv2505}, \href
  {https://ui.adsabs.harvard.edu/abs/2016MNRAS.455.3911D} {455, 3911}

\bibitem[\protect\citeauthoryear{{Elitzur}}{{Elitzur}}{2008}]{elitzur2008}
{Elitzur} M.,  2008, \mn@doi [\nar] {10.1016/j.newar.2008.06.010}, \href
  {https://ui.adsabs.harvard.edu/abs/2008NewAR..52..274E} {52, 274}

\bibitem[\protect\citeauthoryear{{Fanaroff} \& {Riley}}{{Fanaroff} \&
  {Riley}}{1974}]{fanaroff1974}
{Fanaroff} B.~L.,  {Riley} J.~M.,  1974, \mn@doi [\mnras]
  {10.1093/mnras/167.1.31P}, \href
  {https://ui.adsabs.harvard.edu/abs/1974MNRAS.167P..31F} {167, 31P}

\bibitem[\protect\citeauthoryear{{Fanti}, {Fanti}, {de Ruiter}  \&
  {Parma}}{{Fanti} et~al.}{1986}]{fanti1986}
{Fanti} C.,  {Fanti} R.,  {de Ruiter} H.~R.,   {Parma} P.,  1986, \aaps, \href
  {https://ui.adsabs.harvard.edu/abs/1986A&AS...65..145F} {65, 145}

\bibitem[\protect\citeauthoryear{{Fanti}, {Fanti}, {de Ruiter}  \&
  {Parma}}{{Fanti} et~al.}{1987}]{fanti1987}
{Fanti} C.,  {Fanti} R.,  {de Ruiter} H.~R.,   {Parma} P.,  1987, \aaps, \href
  {https://ui.adsabs.harvard.edu/abs/1987A&AS...69...57F} {69, 57}

\bibitem[\protect\citeauthoryear{{Ficarra}, {Grueff}  \&
  {Tomassetti}}{{Ficarra} et~al.}{1985}]{ficarra1985}
{Ficarra} A.,  {Grueff} G.,   {Tomassetti} G.,  1985, \aaps, \href
  {https://ui.adsabs.harvard.edu/abs/1985A&AS...59..255F} {59, 255}

\bibitem[\protect\citeauthoryear{{Gopal-Krishna}, {Wiita}  \&
  {Saripalli}}{{Gopal-Krishna} et~al.}{1989}]{gopal-krishna1989}
{Gopal-Krishna} {Wiita} P.~J.,   {Saripalli} L.,  1989, \mn@doi [\mnras]
  {10.1093/mnras/239.1.173}, \href
  {https://ui.adsabs.harvard.edu/abs/1989MNRAS.239..173G} {239, 173}

\bibitem[\protect\citeauthoryear{{Green}}{{Green}}{2011}]{green2011}
{Green} D.~A.,  2011, \mn@doi [Bulletin of the Astronomical Society of India]
  {10.48550/arXiv.1108.5083}, \href
  {https://ui.adsabs.harvard.edu/abs/2011BASI...39..289G} {39, 289}

\bibitem[\protect\citeauthoryear{{Gregory}, {Scott}, {Douglas}  \&
  {Condon}}{{Gregory} et~al.}{1996}]{gregory1996}
{Gregory} P.~C.,  {Scott} W.~K.,  {Douglas} K.,   {Condon} J.~J.,  1996,
  \mn@doi [\apjs] {10.1086/192282}, \href
  {https://ui.adsabs.harvard.edu/abs/1996ApJS..103..427G} {103, 427}

\bibitem[\protect\citeauthoryear{{G{\"u}rkan} et~al.,}{{G{\"u}rkan}
  et~al.}{2022}]{gurkan2022}
{G{\"u}rkan} G.,  et~al., 2022, \mn@doi [\mnras] {10.1093/mnras/stac880}, \href
  {https://ui.adsabs.harvard.edu/abs/2022MNRAS.512.6104G} {512, 6104}

\bibitem[\protect\citeauthoryear{{Hales}, {Baldwin}  \& {Warner}}{{Hales}
  et~al.}{1988}]{hales1988}
{Hales} S.~E.~G.,  {Baldwin} J.~E.,   {Warner} P.~J.,  1988, \mn@doi [\mnras]
  {10.1093/mnras/234.4.919}, \href
  {https://ui.adsabs.harvard.edu/abs/1988MNRAS.234..919H} {234, 919}

\bibitem[\protect\citeauthoryear{{Hales}, {Masson}, {Warner}  \&
  {Baldwin}}{{Hales} et~al.}{1990}]{hales1990}
{Hales} S.~E.~G.,  {Masson} C.~R.,  {Warner} P.~J.,   {Baldwin} J.~E.,  1990,
  \mnras, \href {https://ui.adsabs.harvard.edu/abs/1990MNRAS.246..256H} {246,
  256}

\bibitem[\protect\citeauthoryear{{Hancock}, {Murphy}, {Gaensler}, {Hopkins}  \&
  {Curran}}{{Hancock} et~al.}{2012}]{hancock2012}
{Hancock} P.~J.,  {Murphy} T.,  {Gaensler} B.~M.,  {Hopkins} A.,   {Curran}
  J.~R.,  2012, \mn@doi [\mnras] {10.1111/j.1365-2966.2012.20768.x}, \href
  {https://ui.adsabs.harvard.edu/abs/2012MNRAS.422.1812H} {422, 1812}

\bibitem[\protect\citeauthoryear{{Hancock}, {Trott}  \&
  {Hurley-Walker}}{{Hancock} et~al.}{2018}]{hancock2018}
{Hancock} P.~J.,  {Trott} C.~M.,   {Hurley-Walker} N.,  2018, \mn@doi [\pasa]
  {10.1017/pasa.2018.3}, \href
  {https://ui.adsabs.harvard.edu/abs/2018PASA...35...11H} {35, e011}

\bibitem[\protect\citeauthoryear{{Hardcastle}}{{Hardcastle}}{2018}]{hardcastle2018}
{Hardcastle} M.~J.,  2018, \mn@doi [\mnras] {10.1093/mnras/stx3358}, \href
  {https://ui.adsabs.harvard.edu/abs/2018MNRAS.475.2768H} {475, 2768}

\bibitem[\protect\citeauthoryear{{Hickish} et~al.,}{{Hickish}
  et~al.}{2018}]{hickish2018}
{Hickish} J.,  et~al., 2018, \mn@doi [\mnras] {10.1093/mnras/sty074}, \href
  {https://ui.adsabs.harvard.edu/abs/2018MNRAS.475.5677H} {475, 5677}

\bibitem[\protect\citeauthoryear{{Hill} et~al.,}{{Hill}
  et~al.}{2008}]{hill2008}
{Hill} G.~J.,  et~al., 2008, in {Kodama} T.,  {Yamada} T.,   {Aoki} K.,  eds,
  Astronomical Society of the Pacific Conference Series Vol. 399, Panoramic
  Views of Galaxy Formation and Evolution. p.~115

\bibitem[\protect\citeauthoryear{{H{\"o}nig} \& {Beckert}}{{H{\"o}nig} \&
  {Beckert}}{2007}]{honig2007}
{H{\"o}nig} S.~F.,  {Beckert} T.,  2007, \mn@doi [\mnras]
  {10.1111/j.1365-2966.2007.12157.x}, \href
  {https://ui.adsabs.harvard.edu/abs/2007MNRAS.380.1172H} {380, 1172}

\bibitem[\protect\citeauthoryear{{Hotan} et~al.,}{{Hotan}
  et~al.}{2021}]{hotan2021}
{Hotan} A.~W.,  et~al., 2021, \mn@doi [\pasa] {10.1017/pasa.2021.1}, \href
  {https://ui.adsabs.harvard.edu/abs/2021PASA...38....9H} {38, e009}

\bibitem[\protect\citeauthoryear{{Intema}, {Jagannathan}, {Mooley}  \&
  {Frail}}{{Intema} et~al.}{2017}]{intema2017}
{Intema} H.~T.,  {Jagannathan} P.,  {Mooley} K.~P.,   {Frail} D.~A.,  2017,
  \mn@doi [\aap] {10.1051/0004-6361/201628536}, \href
  {https://ui.adsabs.harvard.edu/abs/2017A&A...598A..78I} {598, A78}

\bibitem[\protect\citeauthoryear{{Jamrozy}, {Konar}, {Machalski}  \&
  {Saikia}}{{Jamrozy} et~al.}{2008}]{jamrozy2008}
{Jamrozy} M.,  {Konar} C.,  {Machalski} J.,   {Saikia} D.~J.,  2008, \mn@doi
  [\mnras] {10.1111/j.1365-2966.2007.12772.x}, \href
  {https://ui.adsabs.harvard.edu/abs/2008MNRAS.385.1286J} {385, 1286}

\bibitem[\protect\citeauthoryear{{Kaiser} \& {Alexander}}{{Kaiser} \&
  {Alexander}}{1997}]{kaiser1997}
{Kaiser} C.~R.,  {Alexander} P.,  1997, \mn@doi [\mnras]
  {10.1093/mnras/286.1.215}, \href
  {https://ui.adsabs.harvard.edu/abs/1997MNRAS.286..215K} {286, 215}

\bibitem[\protect\citeauthoryear{{Kaiser}, {Dennett-Thorpe}  \&
  {Alexander}}{{Kaiser} et~al.}{1997}]{kaiser1997b}
{Kaiser} C.~R.,  {Dennett-Thorpe} J.,   {Alexander} P.,  1997, \mn@doi [\mnras]
  {10.1093/mnras/292.3.723}, \href
  {https://ui.adsabs.harvard.edu/abs/1997MNRAS.292..723K} {292, 723}

\bibitem[\protect\citeauthoryear{{Kuehr}, {Witzel}, {Pauliny-Toth}  \&
  {Nauber}}{{Kuehr} et~al.}{1981}]{kuehr1981}
{Kuehr} H.,  {Witzel} A.,  {Pauliny-Toth} I.~I.~K.,   {Nauber} U.,  1981,
  \aaps, \href {https://ui.adsabs.harvard.edu/abs/1981A&AS...45..367K} {45,
  367}

\bibitem[\protect\citeauthoryear{{Ku{\'z}micz} \& {Jamrozy}}{{Ku{\'z}micz} \&
  {Jamrozy}}{2012}]{kuzmicz2012}
{Ku{\'z}micz} A.,  {Jamrozy} M.,  2012, \mn@doi [\mnras]
  {10.1111/j.1365-2966.2012.21576.x}, \href
  {https://ui.adsabs.harvard.edu/abs/2012MNRAS.426..851K} {426, 851}

\bibitem[\protect\citeauthoryear{{Lacy} et~al.,}{{Lacy}
  et~al.}{2020}]{lacy2020}
{Lacy} M.,  et~al., 2020, \mn@doi [\pasp] {10.1088/1538-3873/ab63eb}, \href
  {https://ui.adsabs.harvard.edu/abs/2020PASP..132c5001L} {132, 035001}

\bibitem[\protect\citeauthoryear{{Laing}, {Riley}  \& {Longair}}{{Laing}
  et~al.}{1983}]{laing1983}
{Laing} R.~A.,  {Riley} J.~M.,   {Longair} M.~S.,  1983, \mn@doi [\mnras]
  {10.1093/mnras/204.1.151}, \href
  {https://ui.adsabs.harvard.edu/abs/1983MNRAS.204..151L} {204, 151}

\bibitem[\protect\citeauthoryear{{Lane}, {Cotton}, {Helmboldt}  \&
  {Kassim}}{{Lane} et~al.}{2012}]{lane2012}
{Lane} W.~M.,  {Cotton} W.~D.,  {Helmboldt} J.~F.,   {Kassim} N.~E.,  2012,
  \mn@doi [Radio Science] {10.1029/2011RS004941}, \href
  {https://ui.adsabs.harvard.edu/abs/2012RaSc...47.0K04L} {47, RS0K04}

\bibitem[\protect\citeauthoryear{{Lane}, {Cotton}, {van Velzen}, {Clarke},
  {Kassim}, {Helmboldt}, {Lazio}  \& {Cohen}}{{Lane} et~al.}{2014}]{lane2014}
{Lane} W.~M.,  {Cotton} W.~D.,  {van Velzen} S.,  {Clarke} T.~E.,  {Kassim}
  N.~E.,  {Helmboldt} J.~F.,  {Lazio} T.~J.~W.,   {Cohen} A.~S.,  2014, \mn@doi
  [\mnras] {10.1093/mnras/stu256}, \href
  {https://ui.adsabs.harvard.edu/abs/2014MNRAS.440..327L} {440, 327}

\bibitem[\protect\citeauthoryear{{Liu}, {Pooley}  \& {Riley}}{{Liu}
  et~al.}{1992}]{liu1992}
{Liu} R.,  {Pooley} G.,   {Riley} J.~M.,  1992, \mn@doi [\mnras]
  {10.1093/mnras/257.4.545}, \href
  {https://ui.adsabs.harvard.edu/abs/1992MNRAS.257..545L} {257, 545}

\bibitem[\protect\citeauthoryear{{Machalski}}{{Machalski}}{1998}]{machalski1998}
{Machalski} J.,  1998, \mn@doi [\aaps] {10.1051/aas:1998132}, \href
  {https://ui.adsabs.harvard.edu/abs/1998A&AS..128..153M} {128, 153}

\bibitem[\protect\citeauthoryear{{Machalski}, {Chy{\.z}y}, {Stawarz}  \&
  {Kozie{\l}}}{{Machalski} et~al.}{2007}]{machalski2007}
{Machalski} J.,  {Chy{\.z}y} K.~T.,  {Stawarz} {\L}.,   {Kozie{\l}} D.,  2007,
  \mn@doi [\aap] {10.1051/0004-6361:20066121}, \href
  {https://ui.adsabs.harvard.edu/abs/2007A&A...462...43M} {462, 43}

\bibitem[\protect\citeauthoryear{{Machalski}, {Jamrozy}, {Stawarz}  \&
  {Kozie{\l}-Wierzbowska}}{{Machalski} et~al.}{2011}]{machalski2011a}
{Machalski} J.,  {Jamrozy} M.,  {Stawarz} {\L}.,   {Kozie{\l}-Wierzbowska} D.,
  2011, \mn@doi [\apj] {10.1088/0004-637X/740/2/58}, \href
  {https://ui.adsabs.harvard.edu/abs/2011ApJ...740...58M} {740, 58}

\bibitem[\protect\citeauthoryear{{Machalski}, {Kozie{\l}-Wierzbowska}  \&
  {Goyal}}{{Machalski} et~al.}{2021}]{machalski2021}
{Machalski} J.,  {Kozie{\l}-Wierzbowska} D.,   {Goyal} A.,  2021, \mn@doi
  [\apjs] {10.3847/1538-4365/ac08a0}, \href
  {https://ui.adsabs.harvard.edu/abs/2021ApJS..255...22M} {255, 22}

\bibitem[\protect\citeauthoryear{{Mack}, {Klein}, {O'Dea}, {Willis}  \&
  {Saripalli}}{{Mack} et~al.}{1998}]{mack1998}
{Mack} K.~H.,  {Klein} U.,  {O'Dea} C.~P.,  {Willis} A.~G.,   {Saripalli} L.,
  1998, \aap, \href {https://ui.adsabs.harvard.edu/abs/1998A&A...329..431M}
  {329, 431}

\bibitem[\protect\citeauthoryear{{Macklin}}{{Macklin}}{1982}]{macklin1982}
{Macklin} J.~T.,  1982, \mn@doi [\mnras] {10.1093/mnras/199.4.1119}, \href
  {https://ui.adsabs.harvard.edu/abs/1982MNRAS.199.1119M} {199, 1119}

\bibitem[\protect\citeauthoryear{{Nenkova}, {Sirocky}, {Nikutta}, {Ivezi{\'c}}
  \& {Elitzur}}{{Nenkova} et~al.}{2008}]{nenkova2008}
{Nenkova} M.,  {Sirocky} M.~M.,  {Nikutta} R.,  {Ivezi{\'c}} {\v{Z}}.,
  {Elitzur} M.,  2008, \mn@doi [\apj] {10.1086/590483}, \href
  {https://ui.adsabs.harvard.edu/abs/2008ApJ...685..160N} {685, 160}

\bibitem[\protect\citeauthoryear{{O'Sullivan} et~al.,}{{O'Sullivan}
  et~al.}{2019}]{osullivan2019}
{O'Sullivan} S.~P.,  et~al., 2019, \mn@doi [\aap]
  {10.1051/0004-6361/201833832}, \href
  {https://ui.adsabs.harvard.edu/abs/2019A&A...622A..16O} {622, A16}

\bibitem[\protect\citeauthoryear{{Oei} et~al.,}{{Oei} et~al.}{2022}]{oei2022}
{Oei} M. S.~S.~L.,  et~al., 2022, \mn@doi [\aap] {10.1051/0004-6361/202142778},
  \href {https://ui.adsabs.harvard.edu/abs/2022A&A...660A...2O} {660, A2}

\bibitem[\protect\citeauthoryear{{Oei} et~al.,}{{Oei} et~al.}{2023}]{oei2023}
{Oei} M. S.~S.~L.,  et~al., 2023, \mn@doi [\aap] {10.1051/0004-6361/202243572},
  \href {https://ui.adsabs.harvard.edu/abs/2023A&A...672A.163O} {672, A163}

\bibitem[\protect\citeauthoryear{{Oei} et~al.,}{{Oei} et~al.}{2024}]{oei2024}
{Oei} M. S.~S.~L.,  et~al., 2024, \mn@doi [\aap] {10.1051/0004-6361/202347115},
  \href {https://ui.adsabs.harvard.edu/abs/2024A&A...686A.137O} {686, A137}

\bibitem[\protect\citeauthoryear{{Pearson} \& {Kus}}{{Pearson} \&
  {Kus}}{1978}]{pearson1978}
{Pearson} T.~J.,  {Kus} A.~J.,  1978, \mn@doi [\mnras]
  {10.1093/mnras/182.2.273}, \href
  {https://ui.adsabs.harvard.edu/abs/1978MNRAS.182..273P} {182, 273}

\bibitem[\protect\citeauthoryear{{Perley} \& {Butler}}{{Perley} \&
  {Butler}}{2013a}]{perley2013a}
{Perley} R.~A.,  {Butler} B.~J.,  2013a, \mn@doi [\apjs]
  {10.1088/0067-0049/204/2/19}, \href
  {https://ui.adsabs.harvard.edu/abs/2013ApJS..204...19P} {204, 19}

\bibitem[\protect\citeauthoryear{{Perley} \& {Butler}}{{Perley} \&
  {Butler}}{2013b}]{perley2013b}
{Perley} R.~A.,  {Butler} B.~J.,  2013b, \mn@doi [\apjs]
  {10.1088/0067-0049/206/2/16}, \href
  {https://ui.adsabs.harvard.edu/abs/2013ApJS..206...16P} {206, 16}

\bibitem[\protect\citeauthoryear{{Polatidis} \& {Conway}}{{Polatidis} \&
  {Conway}}{2003}]{polatidis2003}
{Polatidis} A.~G.,  {Conway} J.~E.,  2003, \mn@doi [\pasa] {10.1071/AS02053},
  \href {https://ui.adsabs.harvard.edu/abs/2003PASA...20...69P} {20, 69}

\bibitem[\protect\citeauthoryear{{Rawlings}, {Eales}  \& {Lacy}}{{Rawlings}
  et~al.}{2001}]{rawlings2001}
{Rawlings} S.,  {Eales} S.,   {Lacy} M.,  2001, \mn@doi [\mnras]
  {10.1046/j.1365-8711.2001.04151.x}, \href
  {https://ui.adsabs.harvard.edu/abs/2001MNRAS.322..523R} {322, 523}

\bibitem[\protect\citeauthoryear{{Rengelink}, {Tang}, {de Bruyn}, {Miley},
  {Bremer}, {R{\"o}ttgering}  \& {Bremer}}{{Rengelink}
  et~al.}{1997}]{rengelink1997}
{Rengelink} R.~B.,  {Tang} Y.,  {de Bruyn} A.~G.,  {Miley} G.~K.,  {Bremer}
  M.~N.,  {R{\"o}ttgering} H.~J.~A.,   {Bremer} M.~A.~R.,  1997, \mn@doi
  [\aaps] {10.1051/aas:1997358}, \href
  {https://ui.adsabs.harvard.edu/abs/1997A&AS..124..259R} {124, 259}

\bibitem[\protect\citeauthoryear{{Richards} et~al.,}{{Richards}
  et~al.}{2011}]{richards2011}
{Richards} J.~L.,  et~al., 2011, \mn@doi [\apjs] {10.1088/0067-0049/194/2/29},
  \href {https://ui.adsabs.harvard.edu/abs/2011ApJS..194...29R} {194, 29}

\bibitem[\protect\citeauthoryear{{Riley}, {Waldram}  \& {Riley}}{{Riley}
  et~al.}{1999}]{riley1999}
{Riley} J.~M.~W.,  {Waldram} E.~M.,   {Riley} J.~M.,  1999, \mn@doi [\mnras]
  {10.1046/j.1365-8711.1999.02416.x}, \href
  {https://ui.adsabs.harvard.edu/abs/1999MNRAS.306...31R} {306, 31}

\bibitem[\protect\citeauthoryear{{Scheuer}}{{Scheuer}}{1995}]{scheuer1995}
{Scheuer} P.~A.~G.,  1995, \mn@doi [\mnras] {10.1093/mnras/277.1.331}, \href
  {https://ui.adsabs.harvard.edu/abs/1995MNRAS.277..331S} {277, 331}

\bibitem[\protect\citeauthoryear{{Schmitt}, {Donley}, {Antonucci}, {Hutchings},
  {Kinney}  \& {Pringle}}{{Schmitt} et~al.}{2003}]{schmitt2003}
{Schmitt} H.~R.,  {Donley} J.~L.,  {Antonucci} R.~R.~J.,  {Hutchings} J.~B.,
  {Kinney} A.~L.,   {Pringle} J.~E.,  2003, \mn@doi [\apj] {10.1086/381224},
  \href {https://ui.adsabs.harvard.edu/abs/2003ApJ...597..768S} {597, 768}

\bibitem[\protect\citeauthoryear{{Schoenmakers}, {Mack}, {de Bruyn},
  {R{\"o}ttgering}, {Klein}  \& {van der Laan}}{{Schoenmakers}
  et~al.}{2000}]{schoenmakers2000}
{Schoenmakers} A.~P.,  {Mack} K.~H.,  {de Bruyn} A.~G.,  {R{\"o}ttgering}
  H.~J.~A.,  {Klein} U.,   {van der Laan} H.,  2000, \mn@doi [\aaps]
  {10.1051/aas:2000267}, \href
  {https://ui.adsabs.harvard.edu/abs/2000A&AS..146..293S} {146, 293}

\bibitem[\protect\citeauthoryear{{Shimwell} et~al.,}{{Shimwell}
  et~al.}{2019}]{shimwell2019}
{Shimwell} T.~W.,  et~al., 2019, \mn@doi [\aap] {10.1051/0004-6361/201833559},
  \href {https://ui.adsabs.harvard.edu/abs/2019A&A...622A...1S} {622, A1}

\bibitem[\protect\citeauthoryear{{Shimwell} et~al.,}{{Shimwell}
  et~al.}{2022}]{shimwell2022}
{Shimwell} T.~W.,  et~al., 2022, \mn@doi [\aap] {10.1051/0004-6361/202142484},
  \href {https://ui.adsabs.harvard.edu/abs/2022A&A...659A...1S} {659, A1}

\bibitem[\protect\citeauthoryear{{Stuardi} et~al.,}{{Stuardi}
  et~al.}{2020}]{stuardi2020}
{Stuardi} C.,  et~al., 2020, \mn@doi [\aap] {10.1051/0004-6361/202037635},
  \href {https://ui.adsabs.harvard.edu/abs/2020A&A...638A..48S} {638, A48}

\bibitem[\protect\citeauthoryear{{Subrahmanyan}, {Saripalli}  \&
  {Hunstead}}{{Subrahmanyan} et~al.}{1996}]{subrahmanyan1996}
{Subrahmanyan} R.,  {Saripalli} L.,   {Hunstead} R.~W.,  1996, \mn@doi [\mnras]
  {10.1093/mnras/279.1.257}, \href
  {https://ui.adsabs.harvard.edu/abs/1996MNRAS.279..257S} {279, 257}

\bibitem[\protect\citeauthoryear{{Turner}, {Yates-Jones}, {Shabala}, {Quici}
  \& {Stewart}}{{Turner} et~al.}{2023}]{turner2023}
{Turner} R.~J.,  {Yates-Jones} P.~M.,  {Shabala} S.~S.,  {Quici} B.,
  {Stewart} G. S.~C.,  2023, \mn@doi [\mnras] {10.1093/mnras/stac2998}, \href
  {https://ui.adsabs.harvard.edu/abs/2023MNRAS.518..945T} {518, 945}

\bibitem[\protect\citeauthoryear{{Urry} \& {Padovani}}{{Urry} \&
  {Padovani}}{1995}]{urry1995}
{Urry} C.~M.,  {Padovani} P.,  1995, \mn@doi [\pasp] {10.1086/133630}, \href
  {https://ui.adsabs.harvard.edu/abs/1995PASP..107..803U} {107, 803}

\bibitem[\protect\citeauthoryear{{Vikhlinin}, {Kravtsov}, {Forman}, {Jones},
  {Markevitch}, {Murray}  \& {Van Speybroeck}}{{Vikhlinin}
  et~al.}{2006}]{vikhlinin2006}
{Vikhlinin} A.,  {Kravtsov} A.,  {Forman} W.,  {Jones} C.,  {Markevitch} M.,
  {Murray} S.~S.,   {Van Speybroeck} L.,  2006, \mn@doi [\apj]
  {10.1086/500288}, \href
  {https://ui.adsabs.harvard.edu/abs/2006ApJ...640..691V} {640, 691}

\bibitem[\protect\citeauthoryear{{Wiita}, {Rosen}, {Gopal-Krishna,}  \&
  {Saripalli}}{{Wiita} et~al.}{1989}]{wiita1989}
{Wiita} P.~J.,  {Rosen} A.,  {Gopal-Krishna,}  {Saripalli} L.,  1989, in
  {Meisenheimer} K.,  {Roeser} H.-J.,  eds, Springer-Verlag, Berlin, Vol.~327,
  Hot Spots in Extragalactic Radio Sources.
p.~173

\bibitem[\protect\citeauthoryear{{Williams} et~al.,}{{Williams}
  et~al.}{2019}]{williams2019}
{Williams} W.~L.,  et~al., 2019, \mn@doi [\aap] {10.1051/0004-6361/201833564},
  \href {https://ui.adsabs.harvard.edu/abs/2019A&A...622A...2W} {622, A2}

\bibitem[\protect\citeauthoryear{{W{\'o}jtowicz}, {Stawarz}, {Machalski}  \&
  {Ostorero}}{{W{\'o}jtowicz} et~al.}{2021}]{wojtowicz2021}
{W{\'o}jtowicz} A.,  {Stawarz} {\L}.,  {Machalski} J.,   {Ostorero} L.,  2021,
  \mn@doi [\apj] {10.3847/1538-4357/ac116c}, \href
  {https://ui.adsabs.harvard.edu/abs/2021ApJ...922..197W} {922, 197}

\bibitem[\protect\citeauthoryear{{Zwart} et~al.,}{{Zwart}
  et~al.}{2008}]{zwart2008}
{Zwart} J.~T.~L.,  et~al., 2008, \mn@doi [\mnras]
  {10.1111/j.1365-2966.2008.13953.x}, \href
  {https://ui.adsabs.harvard.edu/abs/2008MNRAS.391.1545Z} {391, 1545}

\bibitem[\protect\citeauthoryear{{de Gasperin} et~al.,}{{de Gasperin}
  et~al.}{2023}]{degasperin2023}
{de Gasperin} F.,  et~al., 2023, \mn@doi [\aap] {10.1051/0004-6361/202245389},
  \href {https://ui.adsabs.harvard.edu/abs/2023A&A...673A.165D} {673, A165}

\bibitem[\protect\citeauthoryear{{de Ruiter}, {Parma}, {Fanti}  \& {Fanti}}{{de
  Ruiter} et~al.}{1986}]{deruiter1986}
{de Ruiter} H.~R.,  {Parma} P.,  {Fanti} C.,   {Fanti} R.,  1986, \aaps, \href
  {https://ui.adsabs.harvard.edu/abs/1986A&AS...65..111D} {65, 111}

\makeatother
\end{thebibliography}


\appendix

\section{Radio images and spectra of the GRQ sample}

Radio images of our GRQs at 54~MHz, 144~MHz, 3~GHz and 15.5~GHz are shown in Fig.~\ref{fig:grq_overlays}. The observed radio spectra of the GRQs, along with the best-fit model for the lobe emission used to constrain the dynamical modelling, are presented in Fig.~\ref{fig:grq_spectra}; the observational data points are given in Table~\ref{tab:grq_flux_table}. The measurements of the radio spectra are described in Section~\ref{Radio spectra}. 
 
\begin{figure*}
\begin{center}
\includegraphics[scale=0.9, angle=0, trim=3.1cm 2cm 0cm 3cm]{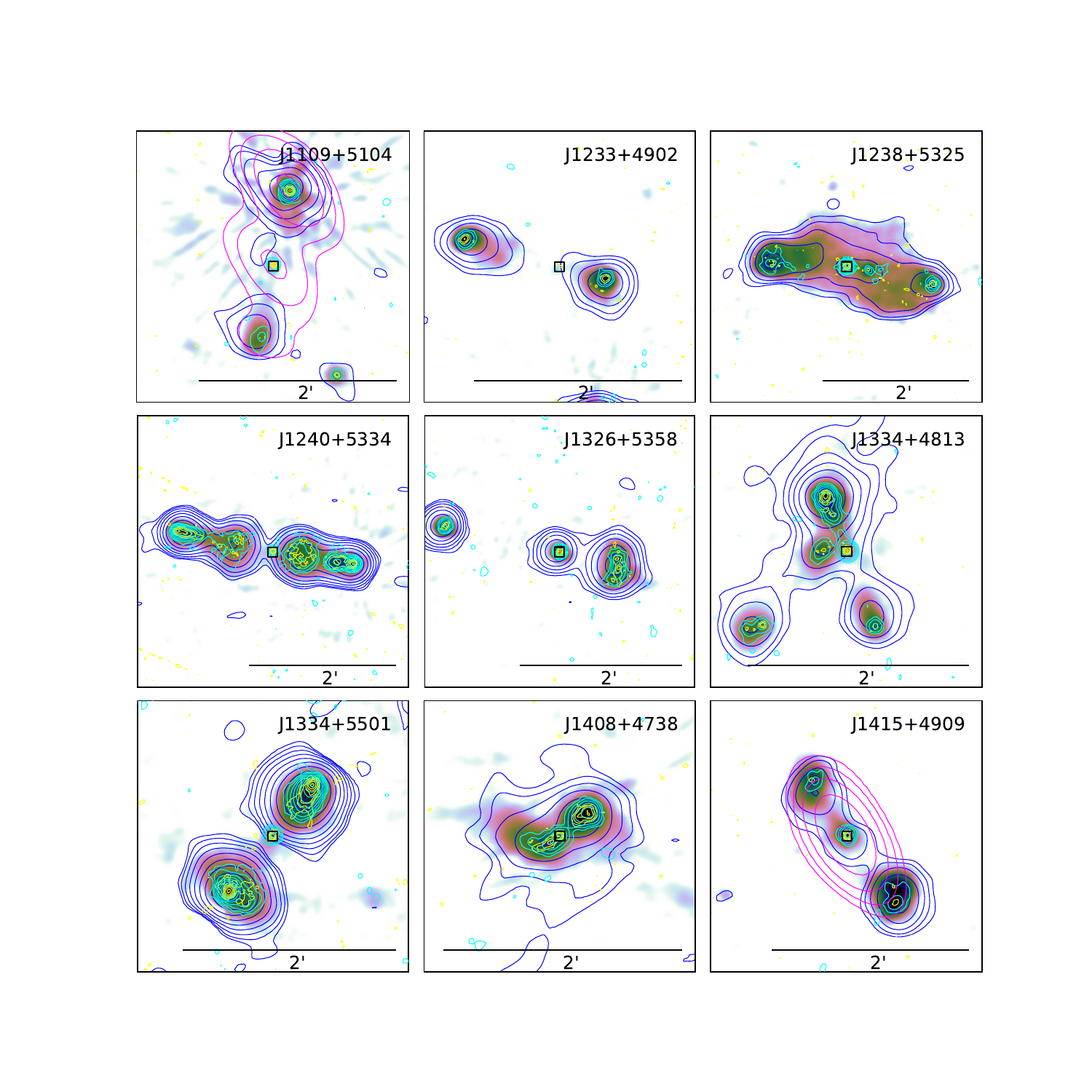}
\caption{Radio contours from LoLSS DR1 (blue), FIRST (cyan) and VLASS (yellow) are overlaid on the LoTSS DR2 images for the GRQs. The LoTSS DR2 images are displayed using the `cubehelix' colour scheme by \protect\cite{green2011}. In addition, for J1109+5104, J1415+4909, J1439+4550 and J1450+4549, radio contours from the AMI images are shown in magenta. For each set of contours, the lowest contour is at the 3$\sigma$ level, where $\sigma$ is the local rms, with the number of $\sigma$ doubling with each subsequent contour. The black square shows the position of the host galaxy. For most sources, the emission from the core and hotspots is detected in the FIRST and VLASS images at 1.4 and 3~GHz, respectively.}
\label{fig:grq_overlays}
\end{center}
\end{figure*}
\FloatBarrier

\setcounter{figure}{0}

\begin{figure*}
\begin{center}
\includegraphics[scale=0.9, angle=0, trim=3.1cm 1.5cm 0cm 2cm]{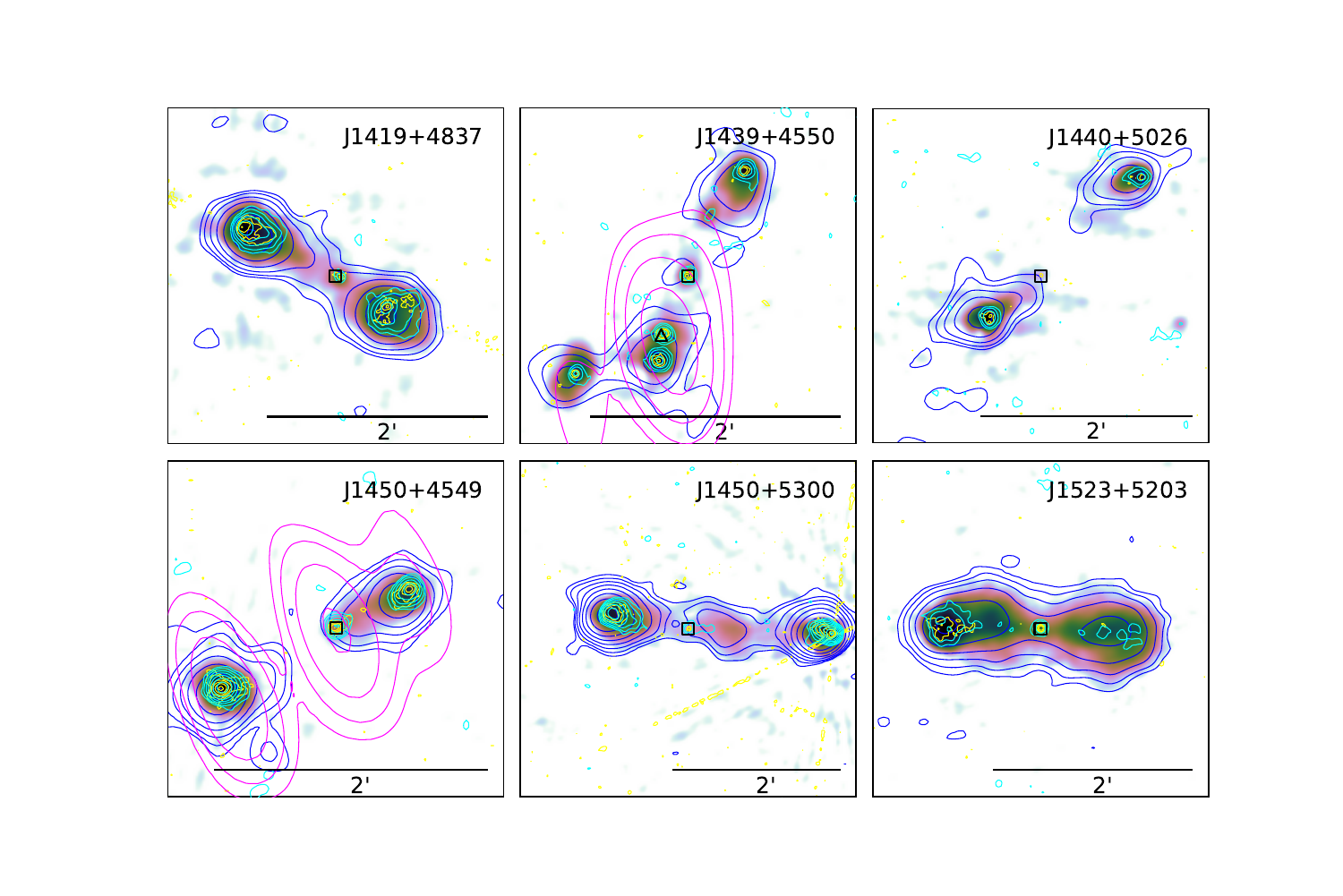}
\caption{Continued. For J1439+4550, the black triangle shows the position of a potentially weak confusing source in the southern lobe. In the VLASS image of J1450+5300, mild (below the $5\sigma$ level) artefacts in the form of spokes are visible around the western lobe.}
\label{fig:grq_overlays}
\end{center}
\end{figure*}

\begin{figure*}
\begin{center}
\includegraphics[scale=0.60, angle=0, trim=1.5cm 2cm 0cm 4cm]{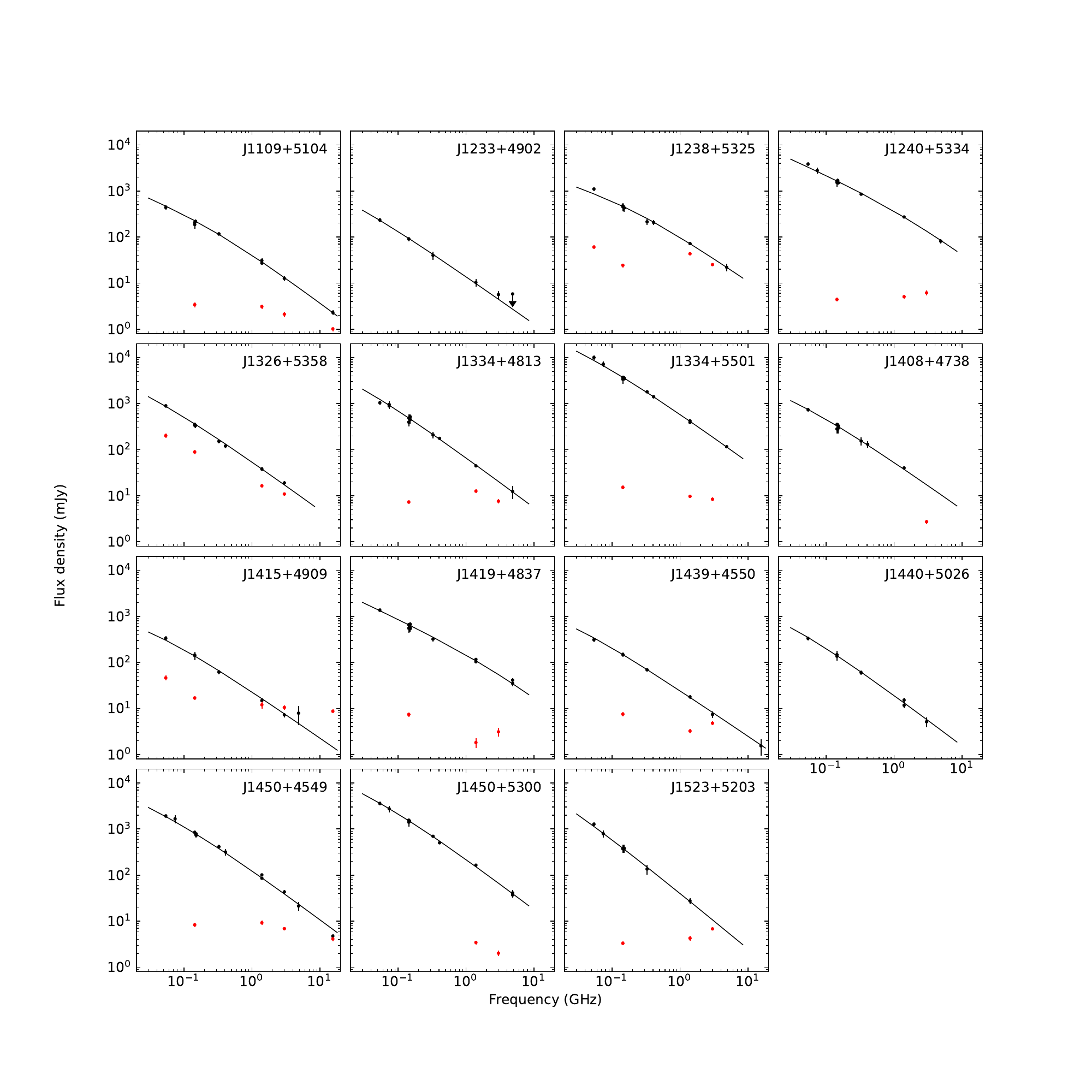}
\caption{Radio spectra of the GRQs. Observed flux densities of the lobes and core are marked with black and red points, respectively. We note that some of the data points are bigger than the error bars. The best-fit model for the lobe emission is drawn with a line.}
\label{fig:grq_spectra}
\end{center}
\end{figure*}

\begin{table}
\caption{Measured flux densities of the GRQs and model fit values. The columns are: (1) source name; (2) observed frequency; (3) lobe flux density; (4) core flux density; (5) reference for the flux density measurements; (6) flux density of the lobes from the fitted model. The core flux densities in brackets are estimated core contributions which are subtracted from the total flux densities to derive the lobe flux densities.}
\label{tab:grq_flux_table}
\begin{tabular}{crccrr}
\hline
Source     & Freq. & $S_{\mathrm{lobes}}$  & $S_{\mathrm{core}}$ & Ref. & Model fit \\
name       & (MHz) &   (mJy)      &   (mJy)  &      &  (mJy) \\
(1)        &  (2)  &    (3)       &   (4)    &  (5) &  (6)   \\     
\hline
J1109+5104 &    54 &  438 $\pm$ 45  &   --     &  (2) & 467.3 \\
           &   144 &  206 $\pm$ 21  & 3.40 $\pm$ 0.40 & (5) & 225.7 \\
           &   144 &  186 $\pm$ 37  &   --     &  (6) & \\
           &   148 &  219 $\pm$ 22  &   --     &  (7) & 218.5 \\
           &   327 &  117 $\pm$ 12  &   --     & (11) & 113.8 \\
           &  1400 & 27.1 $\pm$ 2.2 & 3.09 $\pm$ 0.34 & (14) & 28.7 \\
           &  1400 & 30.8 $\pm$ 3.8 &   --     & (13) & \\    
           &  3000 & 12.6 $\pm$ 1.3 & 2.11 $\pm$ 0.30 & (15) & 13.1 \\
           &  4850 &    --        &   --     &    & 7.92 \\
           & 15500 & 2.30 $\pm$ 0.26& 1.01 $\pm$ 0.11 & (1) & 2.27 \\
\\
J1233+4902 &    54 &  233 $\pm$ 25  &   --     &  (2) & 230.1 \\
           &   144 & 90.3 $\pm$ 9.1 &   --     &  (5) &  93.0 \\
           &   327 &   40 $\pm$ 8   &   --     & (11) &  41.9 \\
           &  1400 & 10.3 $\pm$ 2   &   --     & (13) &  9.72 \\
           &  3000 & 5.60 $\pm$ 1.10&   --     & (15) &  4.45 \\
           &  4850 &  $<$5.8      &   --     & (16) &  2.71 \\
\\
           J1238+5325 &    54 & 1101 $\pm$ 114 & 60.4 $\pm$ 6.0 & (2) & 861.3 \\
           &   144 &  475 $\pm$ 48  & 24.3 $\pm$ 2.3 & (5) & 462.5 \\
           &   144 &  455 $\pm$ 91  &   --     &  (6) & \\
           &   148 &  436 $\pm$ 44  &   --     &  (7) & 453.8 \\
           &   151 &  428 $\pm$ 74  &   --     & (10) & 447.4 \\
           &   327 &  214 $\pm$ 32  &   --     & (11) & 251.7 \\
           &   408 &  207 $\pm$ 26  &   --     & (12) & 210.5 \\
           &  1400 &   72 $\pm$ 4   & 43.2 $\pm$ 1.8 & (13) & 71.4 \\
           &  3000 &     --       & 25.1 $\pm$ 1.2 & (15) & 34.8 \\
           &  4850 &   22 $\pm$ 4   &   (34)   & (16) &  21.9 \\
\\
J1240+5334 &    54 & 3825 $\pm$ 383 &   --     &  (2)   & 3277.3 \\
           &    74 & 2780 $\pm$ 390 &   --     &  (3)   & 2634.0 \\
           &   144 & 1632 $\pm$ 164 & 4.4 $\pm$ 0.4 & (5) & 1642.9 \\
           &   144 & 1527 $\pm$ 305 &   --     &  (6) & \\
           &   148 & 1690 $\pm$ 170 &   --     &  (7) & 1610.3 \\
           &   151 & 1497 $\pm$ 108 &   --     & (10) & 1587.5 \\
           &   327 &  849 $\pm$ 46  &   --     & (11) &  891.7 \\
           &  1400 &  272 $\pm$ 10  &5.05 $\pm$ 0.45& (13) & 269.0 \\
           &  3000 &     --       &6.13 $\pm$ 0.75& (15) & 133.9 \\
           &  4850 &   81 $\pm$ 8   &  (7.0)   & (16) &   84.4 \\
\\
J1326+5358 &    54 &  897 $\pm$ 90  &  202 $\pm$ 21 &  (2) & 871.1 \\
           &   144 &  338 $\pm$ 34  &   89 $\pm$ 9  &  (5) & 362.4 \\
           &   144 &  354 $\pm$ 36  &    (90)     &  (6) & \\
           &   148 &  335 $\pm$ 40  &     --      &  (7) & 353.3 \\
           &   327 &  152 $\pm$ 16  &    (36)     & (11) & 165.1 \\
           &   408 &  120 $\pm$ 13  &    (33)     & (12) & 132.7 \\
           &  1400 &   38 $\pm$ 4   & 16.4 $\pm$ 1.4 & (13) & 38.0 \\
           &  3000 & 19.0 $\pm$ 2.0 & 10.9 $\pm$ 0.8 & (15) & 17.2 \\
           &  4850 &     --       &      --       &      & 10.4 \\
\\
J1334+4813 &    54 & 1045 $\pm$ 106 &     --    &  (2) & 1233.0 \\
           &    74 &  890 $\pm$ 130 &     --    &  (3) &  922.1 \\
           &    74 &  960 $\pm$ 150 &     --    &  (4) & \\
           &   144 &  396 $\pm$ 80  & 7.3 $\pm$ 0.7 & (5) & 486.1 \\
           &   144 &  487 $\pm$ 97  &     --    &  (6) & \\
           &   148 &  463 $\pm$ 50  &     --    &  (7) & 473.1 \\           
           &   151 &  450 $\pm$ 80  &     --    &  (8) & 463.8 \\
\end{tabular}
\end{table}
 
\setcounter{table}{0}

\begin{table}
\caption{Continued}
\begin{tabular}{crccrr}
\hline
Source     & Freq. & $S_{\mathrm{lobes}}$ & $S_{\mathrm{core}}$ & Ref. & Model fit \\
name       & (MHz) &   (mJy)      &   (mJy)  &      &  (mJy) \\
(1)        &  (2)  &    (3)       &   (4)    &  (5) &  (6)   \\     
\hline
           &   151 &  518 $\pm$ 66  &     --    & (10) & \\
           &   327 &  209 $\pm$ 32  &     --    & (11) & 212.7 \\
           &   408 &  176 $\pm$ 10  &   (9.0)   & (12) & 169.3 \\
           &  1400 & 44.5 $\pm$ 3   & 12.6 $\pm$ 1.2 & (13) & 46.2 \\
           &  3000 &     --       &  7.6 $\pm$ 0.8 & (15) & 20.4 \\
           &  4850 & 12.4 $\pm$ 4   &   (9.0)   & (16) &  12.1 \\
\\
J1334+5501 &    54 &10033 $\pm$ 1005&    (15)   &  (2) & 8624.4 \\
           &    74 & 7050 $\pm$ 860 &     --    &  (3) & 6642.4 \\
           &    74 & 7240 $\pm$ 940 &     --    &  (4) & \\
           &   144 & 3709 $\pm$ 371 & 15.2 $\pm$ 1.5 & (5) & 3718.1 \\
           &   144 & 3380 $\pm$ 676 &     --    &  (6) & \\
           &   148 & 3678 $\pm$ 368 &     --    &  (7) & 3627.7 \\
           &   151 & 3550 $\pm$ 150 &     --    &  (9) & 3563.1 \\
           &   151 & 3456 $\pm$ 240 &     --    & (10) & \\
           &   327 & 1803 $\pm$ 70  &    (14)   & (11) & 1737.1 \\
           &   408 & 1407 $\pm$ 75  &    (13)   & (12) & 1404.5 \\
           &  1400 &  392 $\pm$ 20  & 9.7 $\pm$ 0.6 & (14) & 411.9\\
           &  1400 &  422 $\pm$ 24  &    (10)   & (13) & \\
           &  3000 &     --       & 8.4 $\pm$ 0.8 & (15) & 188.6 \\
           &  4850 &  116 $\pm$ 10  &   (8.0)   & (16) &  114.2 \\
\\
J1408+4738 &    54 &  742 $\pm$ 75  &     --    &  (2) & 748.5 \\
           &   144 &  350 $\pm$ 35  &     --    &  (5) & 329.3 \\
           &   144 &  282 $\pm$ 56  &     --    &  (6) & \\
           &   148 &  278 $\pm$ 29  &     --    &  (7) & 321.3 \\
           &   151 &  300 $\pm$ 77  &     --    &  (8) & 315.6 \\
           &   151 &  329 $\pm$ 44  &     --    & (10) & \\
           &   327 &  154 $\pm$ 31  &     --    & (11) & 155.1 \\
           &   408 &  131 $\pm$ 21  &     --    & (12) & 125.6 \\
           &  1400 & 40.1 $\pm$ 2.8 &   (2.7)   & (13) &  37.4 \\
           &  3000 &     --       & 2.72 $\pm$ 0.3 & (15) & 17.3 \\
           &  4850 &     --       &     --    &      &  10.6 \\
\\
J1415+4909 &    54 &  333 $\pm$ 34  & 46.0 $\pm$ 5.6 &  (2) & 302.1 \\
           &   144 &  144 $\pm$ 18  & 16.9 $\pm$ 1.7 &  (5) & 137.2 \\
           &   144 &  141 $\pm$ 28  &     --    &  (6) & \\
           &   327 & 61.5 $\pm$ 7.0 &     --    &  (1) &    65.7 \\
           &  1400 & 14.8 $\pm$ 2.1 & 12.0 $\pm$ 2.1 & (14) & 16.3 \\
           & 1400 & 12.0 $\pm$ 2.0 & -- & (13) & \\
           &  3000 & 7.23 $\pm$ 0.8 & 10.5 $\pm$ 1.1 & (15) & 7.63 \\
           &  4850 &  7.9 $\pm$ 3.5 &   (10.1)  & (16) &    4.71 \\
           & 15500 &     --       & 8.69 $\pm$ 0.87 & (1) & 1.45 \\
\\
J1419+4837 &    54 & 1357 $\pm$ 136 &     --    &  (2) & 1323.5 \\
           &   144 &  659 $\pm$ 66  &  7.4 $\pm$ 0.8 & (5)& 651.8 \\
           &   144 &  552 $\pm$ 110 &     --    &  (6) & \\
           &   148 &  590 $\pm$ 60  &     --    &  (7) & 638.9 \\
           &   151 &  670 $\pm$ 82  &     --    &  (8) & 629.4 \\
           &   151 &  540 $\pm$ 79  &     --    &  (9) & \\
           &   151 &  560 $\pm$ 59  &     --    & (10) & \\
           &   327 &  320 $\pm$ 34  &     --    & (11) & 352.2 \\
           &  1400 &  116 $\pm$ 6   &    (4.0)  & (13) & 108.0 \\
           &  1400 &  103 $\pm$ 5   & 1.82 $\pm$ 0.45 & (14) & \\
           &  3000 &     --       & 3.10 $\pm$ 0.68 & (15) &54.4 \\
           &  4850 &   41 $\pm$ 5   &    (2.0)  & (16) &  34.4 \\
           &  4850 &   35 $\pm$ 5   &    (2.0)  & (17) & \\
\\
J1439+4550 &    54 &  310 $\pm$ 32  &     --    &  (2) & 339.5 \\
           &   144 &  148 $\pm$ 15  & 7.50 $\pm$ 0.80 & (5) & 147.5 \\
           &   327 & 68.8 $\pm$ 3.5 &    (5.2)  & (11) &  69.7 \\
           &  1400 & 17.8 $\pm$ 1.6 & 3.23 $\pm$ 0.35 & (13) & 17.3 \\
           &  3000 & 7.39 $\pm$ 1.14& 4.78 $\pm$ 0.48 & (15) & 8.19 \\
\end{tabular}
\end{table}
 
\setcounter{table}{0}

\begin{table}
\caption{Continued}
\begin{tabular}{crccrr}
\hline
Source     & Freq. & $S_{\mathrm{lobes}}$ & $S_{\mathrm{core}}$ & Ref. & Model fit \\
name       & (MHz) &   (mJy)      &   (mJy)  &      &  (mJy) \\
(1)        &  (2)  &    (3)       &   (4)    &  (5) &  (6)   \\     
\hline
           &  4850 &     --       &    --     &      & 5.09 \\
           & 15500 & 1.54 $\pm$ 0.6 &    (4.8)  & (1) & 1.60 \\
J1440+5026 &    54 &  332 $\pm$ 34  &     --    &  (2) & 349.1 \\
           &   144 &  147 $\pm$ 30  &     --    &  (5) & 139.8 \\
           &   144 &  136 $\pm$ 27  &     --    &  (6) & \\
           &   327 &   60 $\pm$ 6   &     --    & (11) &  61.4 \\
           &  1400 & 15.2 $\pm$ 2.0 &     --    & (13) &  13.2 \\
           &  1400 & 11.7 $\pm$ 1.7 &     --    & (14) & \\
           &  3000 & 5.13 $\pm$ 1.23&     --    & (15) &  5.77 \\
           &  4850 &     --       &     --    &      &  3.41 \\
\\
J1450+4549 &    54 & 1914 $\pm$ 194 &     --    &  (2) & 1854.9 \\
           &    74 & 1660 $\pm$ 320 &     --    &  (3) & 1433.9 \\
           &   144 &  844 $\pm$ 86  & 8.30 $\pm$ 0.84 & (5) & 805.6 \\
           &   148 &  788 $\pm$ 81  &     --    &  (7) & 785.8 \\
           &   151 &  730 $\pm$ 84  &     --    &  (8) & 772.0 \\
           &   151 &  785 $\pm$ 53  &     --    & (10) & \\
           &   327 &  413 $\pm$ 43  &     --    & (11) & 374.8 \\
           &   408 &  317 $\pm$ 53  &     --    & (12) & 302.2 \\
           &  1400 &  101 $\pm$ 5   &   (9.2)   & (13) &  86.7 \\
           &  1400 & 85.4 $\pm$ 6   & 9.22 $\pm$ 0.93 & (14) & \\
           &  3000 & 43.1 $\pm$ 4.3 & 6.86 $\pm$ 0.65 & (15) & 38.9 \\
           &  4850 & 21.3 $\pm$ 4.5 &   (6.1)   & (16) &  23.3 \\
           & 15500 & 4.73 $\pm$ 0.53& 4.10 $\pm$ 0.43 & (1) & 6.63 \\
\\             
J1450+5300 &    54 & 3576 $\pm$ 358 &     --    &  (2) & 3611.8 \\
           &    74 & 2750 $\pm$ 460 &     --    &  (3) & 2758.2 \\
           &   144 & 1551 $\pm$ 155 &     --    &  (5) & 1509.8 \\
           &   144 & 1407 $\pm$ 280 &     --    &  (6) & \\
           &   148 & 1482 $\pm$ 148 &     --    &  (7) & 1471.4 \\
           &   327 &  696 $\pm$ 46  &     --    & (11) &  682.5 \\
           &   408 &  502 $\pm$ 34  &     --    & (12) &  546.4 \\
           &  1400 &  164 $\pm$ 10  & 3.43 $\pm$ 0.36 & (14) & 151.3 \\
           &  1400 &  149 $\pm$ 7 & -- & (13) & \\
           &  3000 &     --       & 2.01 $\pm$ 0.26 & (15) &  66.6 \\
           &  4850 &   37 $\pm$ 5   &   (2.0)   & (16) &  39.4 \\
           &  4850 &   41 $\pm$ 7   &   (2.0)   & (17) & \\
\\
J1523+5203 &    54 & 1272 $\pm$ 128 &     --    &  (2) & 1134.0 \\
           &    74 &  790 $\pm$ 130 &     --    &  (3) &  803.7 \\
           &   144 &  372 $\pm$ 38  &  3.3 $\pm$ 0.3 &(5) & 381.3 \\
           &   144 &  382 $\pm$ 77  &     --    &  (6) & \\
           &   151 &  380 $\pm$ 80  &     --    &  (9) &  361.3 \\
           &   327 &  135 $\pm$ 32  &     --    & (11) &  148.3 \\
           &  1400 & 27.4 $\pm$ 4.3 & 4.26 $\pm$ 0.48 & (13) & 26.6 \\
           &  3000 &     --       & 6.79 $\pm$ 0.23 & (15) &10.65 \\
           &  4850 &     --       &     --    &      &  5.96 \\
\hline
\end{tabular}
\newline
\textbf{References:} (1) this paper; (2) LoLSS DR1 \citep{degasperin2023}; (3) VLSSr \citep{lane2014}; (4) VLSS \citep{cohen2007}; (5) LoTSS DR2 \citep{shimwell2022}; (6) LoTSS DR1 \citep{dabhade2020}; (7) TGSS ADR1 \citep{intema2017}; (8) 6C2 \citep{hales1988}; (9) 6C3 \citep{hales1990}; (10) 7C \citep{riley1999}; (11) WENSS \citep{rengelink1997}; (12) B3.3 \citep{ficarra1985}; (13) NVSS \citep{condon1998}; (14) FIRST \citep{becker1995}; (15) VLASS \citep{lacy2020}; (16) GB6 \citep{gregory1996}; (17) \cite{becker1991}.
\end{table}

\FloatBarrier


\section{Dynamical modelling predictions for a range of $\beta$ values}\label{Refitting}

In the dynamical modelling of our GRQs, a power-law density profile with $\beta = 1.5$ is assumed (see Section~\ref{Dynamical modelling}). Here, we repeat the modelling procedure for $\beta$ values of 0.5, 1.0 and 1.9, and investigate the sensitivity of the fitted model parameters to the assumed value of $\beta$.

Table~\ref{tab:refitted_model_parameters} shows that $\rho_{0}$ varies by a factor of $\approx 50-200$ for $\beta$ values ranging between 0.5 and 1.9. The remaining parameters are only weakly dependent on $\beta$: for each source, we calculate the ratio $R_{\rho_{\rm X}} = \rho_{\rm X, 1.5}/\rho_{\rm X, mean}$, where $\rho_{\rm X, 1.5}$ is the value of $\rho_{\rm X}$ for $\beta = 1.5$ and $\rho_{\rm X, mean}$ is the mean value of $\rho_{\rm X}$ for all values of $\beta$. The mean value of $R_{\rho_{\rm X}}$ ($0.81 \pm 0.14$) is close to unity, indicating that the values of $\rho_{\rm X}$ determined with $\beta=1.5$ are reasonably close to those determined with $\beta$ values ranging between 0.5--1.9 and can be reliably used in this statistical study. The mean values of $R_{t_{\ell}}$ ($1.03 \pm 0.04$), $R_{Q_{\rm j}}$ ($0.94 \pm 0.07$), $R_{\alpha_{\rm em}}$ ($1.00 \pm 0.002$), $R_{\rho_{v_{\rm h}/c}}$ ($1.01 \pm 0.07$) and $R_{B_{\ell}}$ ($1.01 \pm 0.02$) are also close to unity.

\begin{table*}
\caption{Results of the dynamical modelling for all of our GRQs assuming $\beta$ values of 0.5, 1.0, 1.5 and 1.9. For each source, the results for $\beta = 1.5$ in the third row repeat columns (2)--(5), (7)--(9) and (11) from Table~\ref{tab:model_parameters}, for ease of comparison. In the first, second and fourth rows showing the results for the other values of $\beta$, $\alpha_{\rm inj}$ is fixed to the best-fitted value of $\alpha_{\rm inj}$ for $\beta = 1.5$. If the lowest reduced $\chi^{2}$-value from the model fitting, listed in column (4), is obtained for $\beta \neq 1.5$, a new best-fitted value of $\alpha_{\rm inj}$ is found for this value of $\beta$; the results using the new value of $\alpha_{\rm inj}$ are shown in the fifth row. The mean $\pm$ standard deviation of $t_{\ell}$, $Q_{\rm j}$, $\rho_{\rm X}$, $\alpha_{\rm em}$, $v_{\rm h}$/c and $B_{\ell}$ are shown in the last row.}
\setlength{\tabcolsep}{4pt}
\begin{tabular}{ccccccccccccc}
\hline
Source name & $\beta$ & $\alpha_{\rm inj}$ & $\chi^{2}_{\rm red}$ & $t_{\ell}$ & $Q_{\rm j}$ & $\rho_{0}$ & $\rho_{\rm X}$ & $\alpha_{\rm em}$ & $v_{\rm h}/c$ & $B_{\ell}$ \\
&  & & & (Myr) & ($10^{38}$\,W) & ($10^{-24}$\,kg\,m$^{-3}$) & ($10^{-26}$\,kg\,m$^{-3}$) & & & (nT) \\
(1) & (2) & (3) & (4) & (5) & (6) & (7) & (8) & (9) & (10) & (11) \\
\hline
J1109+5104 & 0.5 & 0.62 & 0.378 & 13.8 & 4.22 & 0.131 & 2.09 & 0.947 & 0.062 & 0.66 \\
 & 1.0 & 0.62 & 0.365 & 14.6 & 3.69 & 0.56 & 1.43 & 0.937 & 0.066 & 0.70 \\
 & 1.5 & 0.62 & 0.469 & 16.5 & 3.67 & 2.03 & 0.694 & 0.939 & 0.075 & 0.67 \\
 & 1.9 & 0.62 & 0.523 & 18.6 & 2.72 & 8.79 & 0.850 & 0.939 & 0.066 & 0.77 \\
 & 1.0 & 0.579 & 0.293 & 19.4 & 2.46 & 8.74 & 2.23 & 0.944 & 0.049 & 0.65 \\
 & Mean & & & $16.6 \pm 2.2$ & $3.35 \pm 0.66$ & & $1.46 \pm 0.62$ & $0.941 \pm 0.0037$ & $0.0636 \pm 0.0085$ & $0.690 \pm 0.043$ \\ 
 \\
J1233+4902 & 0.5 & 0.547 & 0.794 & 32.0 & 1.53 & 0.130 & 2.03 & 1.009 & 0.028 & 0.48 \\
 & 1.0 & 0.547 & 0.747 & 37.4 & 1.32 & 0.734 & 1.79 & 1.007 & 0.027 & 0.50 \\
 & 1.5 & 0.547 & 0.668 & 40.0 & 1.21 & 3.14 & 1.20 & 1.000 & 0.029 & 0.51 \\
 & 1.9 & 0.547 & 0.589 & 42.0 & 1.14 & 9.38 & 0.81 & 0.994 & 0.031 & 0.52 \\
 & 1.9 & 0.538 & 0.535 & 44.9 & 1.07 & 10.8 & 0.93 & 0.988 & 0.029 & 0.52 \\
 & Mean & & & $39.3 \pm 4.4$ & $1.25 \pm 0.16$ & & $1.35 \pm 0.48$ & $1.000 \pm 0.0079$ & $0.0288 \pm 0.0013$ & $0.506 \pm 0.015$ \\ 
 \\
J1238+5325 & 0.5 & 0.530 & 0.407 & 78.8 & 0.194 & 0.060 & 1.34 & 0.913 & 0.011 & 0.21 \\
 & 1.0 & 0.530 & 0.352 & 84.0 & 0.185 & 0.278 & 0.695 & 0.904 & 0.012 & 0.21 \\
 & 1.5 & 0.530 & 0.301 & 91.0 & 0.176 & 1.26 & 0.499 & 0.894 & 0.012 & 0.21 \\
 & 1.9 & 0.530 & 0.291 & 104 & 0.161 & 4.68 & 0.423 & 0.894 & 0.012 & 0.22 \\
 & 1.9 & 0.640 & 0.227 & 52.0 & 0.456 & 1.65 & 0.149 & 0.889 & 0.024 & 0.26 \\
 & Mean & & & $82.0 \pm 17$ & $0.234 \pm 0.11$ & & $0.621 \pm 0.40$ & $0.899 \pm 0.0086$ & $0.0142 \pm 0.0049$ & $0.222 \pm 0.019$ \\ 
 \\
J1240+5334 & 0.5 & 0.640 & 0.923 & 40.0 & 0.695 & 0.258 & 4.32 & 0.872 & 0.019 & 0.46 \\
 & 1.0 & 0.640 & 0.811 & 41.0 & 0.657 & 1.00 & 2.83 & 0.866 & 0.021 & 0.47 \\
 & 1.5 & 0.640 & 0.737 & 43.4 & 0.616 & 4.01 & 1.88 & 0.863 & 0.023 & 0.49 \\
 & 1.9 & 0.640 & 0.763 & 46.0 & 0.574 & 11.6 & 1.30 & 0.860 & 0.024 & 0.50 \\
 & Mean & & & $42.6 \pm 2.3$ & $0.635 \pm 0.045$ & & $2.58 \pm 1.1$ & $0.865 \pm 0.0044$ & $0.0217 \pm 0.0019$ & $0.480 \pm 0.016$ \\ 
 \\
J1326+5358 & 0.5 & 0.580 & 0.813 & 160 & 0.146 & 3.41 & 53.2 & 1.027 & 0.0056 & 0.37 \\
 & 1.0 & 0.580 & 0.722 & 184 & 0.124 & 18.5 & 44.5 & 1.024 & 0.0055 & 0.39 \\
 & 1.5 & 0.580 & 0.659 & 213 & 0.106 & 95.9 & 36.2 & 1.020 & 0.0054 & 0.41 \\
 & 1.9 & 0.580 & 0.636 & 249 & 0.092 & 37.0 & 31.5 & 1.018 & 0.0052 & 0.43 \\
 & 1.9 & 0.603 & 0.607 & 198 & 0.116 & 236 & 20.1 & 1.027 & 0.0066 & 0.43 \\
 & Mean & & & $201 \pm 30$ & $0.117 \pm 0.018$ & & $37.1 \pm 11$ & $1.02 \pm 0.0037$ & $0.00566 \pm 0.00049$ & $0.406 \pm 0.023$ \\ 
 \\
J1334+4813 & 0.5 & 0.604 & 0.922 & 11.6 & 23.2 & 0.148 & 2.37 & 1.054 & 0.073 & 1.25 \\
 & 1.0 & 0.604 & 0.883 & 13.4 & 20.0 & 0.795 & 2.03 & 1.050 & 0.072 & 1.31 \\
 & 1.5 & 0.604 & 0.837 & 15.4 & 17.4 & 3.95 & 1.61 & 1.046 & 0.071 & 1.36 \\
 & 1.9 & 0.604 & 0.824 & 18.0 & 15.2 & 14.9 & 1.40 & 1.044 & 0.069 & 1.42 \\
 & 1.9 & 0.630 & 0.811 & 14.0 & 19.8 & 9.15 & 0.86 & 1.053 & 0.088 & 1.43 \\
 & Mean & & & $14.5 \pm 2.1$ & $19.1 \pm 2.7$ & & $1.65 \pm 0.52$ & $1.05 \pm 0.0039$ & $0.0746 \pm 0.0068$ & $1.35 \pm 0.068$ \\ 
 \\
J1334+5501 & 0.5 & 0.580 & 0.563 & 24.5 & 12.9 & 0.314 & 5.08 & 0.986 & 0.034 & 1.19 \\
 & 1.0 & 0.580 & 0.708 & 27.0 & 12.0 & 1.54 & 4.04 & 0.979 & 0.035 & 1.23 \\
 & 1.5 & 0.580 & 0.483 & 33.5 & 10.3 & 9.24 & 3.91 & 0.983 & 0.032 & 1.28 \\
 & 1.9 & 0.580 & 0.503 & 37.8 & 9.56 & 32.7 & 3.23 & 0.979 & 0.032 & 1.32 \\
 & Mean & & & $30.7 \pm 5.3$ & $11.2 \pm 1.3$ & & $4.07 \pm 0.66$ & $0.982 \pm 0.0029$ & $0.0333 \pm 0.0013$ & $1.26 \pm 0.049$ \\ 
 \\
J1408+4738 & 0.5 & 0.554 & 0.919 & 20.2 & 3.77 & 0.166 & 2.81 & 0.981 & 0.038 & 0.76 \\
 & 1.0 & 0.554 & 0.741 & 20.7 & 3.55 & 0.638 & 1.83 & 0.968 & 0.041 & 0.77 \\
 & 1.5 & 0.554 & 0.739 & 24.9 & 3.03 & 3.35 & 1.62 & 0.968 & 0.039 & 0.81 \\
 & 1.9 & 0.554 & 0.769 & 29.8 & 2.62 & 12.7 & 1.49 & 0.969 & 0.037 & 0.84 \\
 & Mean & & & $23.9 \pm 3.9$ & $3.24 \pm 0.45$ & & $1.94 \pm 0.52$ & $0.971 \pm 0.0055$ & $0.0387 \pm 0.0015$ & $0.795 \pm 0.032$ \\
\label{tab:refitted_model_parameters}
\end{tabular}
\end{table*}

\setcounter{table}{0}

\begin{table*}
\caption{Continued}
\setlength{\tabcolsep}{4pt}
\begin{tabular}{ccccccccccccc}
\hline
Source name & $\beta$ & $\alpha_{\rm inj}$ & $\chi^{2}_{\rm red}$ & $t_{\ell}$ & $Q_{\rm j}$ & $\rho_{0}$ & $\rho_{\rm X}$ & $\alpha_{\rm em}$ & $v_{\rm h}/c$ & $B_{\ell}$ \\
&  & & & (Myr) & ($10^{38}$\,W) & ($10^{-24}$\,kg\,m$^{-3}$) & ($10^{-26}$\,kg\,m$^{-3}$) & & & (nT) \\
(1) & (2) & (3) & (4) & (5) & (6) & (7) & (8) & (9) & (10) & (11) \\
\hline
J1415+4909 & 0.5 & 0.535 & 0.568 & 18.6 & 2.98 & 0.057 & 0.85 & 0.966 & 0.053 & 0.48 \\
 & 1.0 & 0.535 & 0.740 & 19.3 & 2.71 & 0.253 & 0.56 & 0.954 & 0.057 & 0.49 \\
 & 1.5 & 0.535 & 0.735 & 22.4 & 2.31 & 1.37 & 0.45 & 0.950 & 0.056 & 0.51 \\
 & 1.9 & 0.535 & 0.884 & 25.0 & 2.08 & 4.90 & 0.35 & 0.945 & 0.057 & 0.53 \\
 & 0.5 & 0.544 & 0.557 & 18.8 & 3.10 & 0.054 & 0.80 & 0.972 & 0.054 & 0.48 \\
 & Mean & & & $20.8 \pm 2.5$ & $2.64 \pm 0.39$ & & $0.602 \pm 0.19$ & $0.957 \pm 0.010$ & $0.0554 \pm 0.0016$ & $0.498 \pm 0.019$ \\ 
 \\
J1419+4837 & 0.5 & 0.680 & 1.230 & 19.0 & 2.24 & 0.0706 & 1.148 & 0.847 & 0.043 & 0.53 \\
 & 1.0 & 0.680 & 1.156 & 19.0 & 2.15 & 0.267 & 0.710 & 0.845 & 0.048 & 0.54 \\
 & 1.5 & 0.680 & 1.100 & 19.2 & 2.07 & 0.975 & 0.424 & 0.841 & 0.055 & 0.56 \\
 & 1.9 & 0.680 & 1.131 & 20.4 & 1.94 & 2.90 & 0.296 & 0.843 & 0.058 & 0.57 \\
 & Mean & & & $19.4 \pm 0.58$ & $2.10 \pm 0.11$ & & $0.644 \pm 0.33$ & $0.844 \pm 0.0022$ & $0.0510 \pm 0.0059$ & $0.550 \pm 0.016$ \\ 
 \\
J1439+4550 & 0.5 & 0.510 & 1.484 & 12.8 & 7.03 & 0.0218 & 0.330 & 0.959 & 0.074 & 0.56 \\
 & 1.0 & 0.510 & 1.137 & 13.6 & 6.48 & 0.101 & 0.233 & 0.952 & 0.078 & 0.57 \\
 & 1.5 & 0.510 & 0.512 & 14.8 & 5.81 & 0.460 & 0.161 & 0.945 & 0.082 & 0.58 \\
 & 1.9 & 0.510 & 0.932 & 19.4 & 4.86 & 2.45 & 0.189 & 0.950 & 0.071 & 0.61 \\
 & Mean & & & $15.2 \pm 2.6$ & $6.04 \pm 0.81$ & & $0.228 \pm 0.064$ & $0.952 \pm 0.0050$ & $0.0762 \pm 0.0041$ & $0.580 \pm 0.019$ \\ 
 \\
J1440+5026 & 0.5 & 0.622 & 0.677 & 15.7 & 9.24 & 0.034 & 0.446 & 1.064 & 0.083 & 0.52 \\
 & 1.0 & 0.622 & 0.606 & 16.5 & 8.27 & 0.179 & 0.301 & 1.054 & 0.088 & 0.54 \\
 & 1.5 & 0.622 & 0.606 & 17.9 & 7.30 & 0.947 & 0.206 & 1.045 & 0.093 & 0.56 \\
 & 1.9 & 0.622 & 0.603 & 19.7 & 6.50 & 3.62 & 0.154 & 1.039 & 0.095 & 0.57 \\
 & 1.9 & 0.628 & 0.588 & 19.4 & 6.76 & 3.60 & 0.153 & 1.043 & 0.097 & 0.58 \\
 & Mean & & & $17.8 \pm 1.6$ & $7.61 \pm 1.0$ & & $0.252 \pm 0.11$ & $1.05 \pm 0.0090$ & $0.0912 \pm 0.0051$ & $0.554 \pm 0.022$ \\ 
 \\
J1450+4549 & 0.5 & 0.610 & 1.552 & 13.1 & 15.7 & 0.368 & 5.80 & 1.006 & 0.067 & 1.19 \\
 & 1.0 & 0.610 & 1.525 & 14.3 & 13.5 & 1.72 & 4.28 & 0.998 & 0.069 & 1.26 \\
 & 1.5 & 0.610 & 1.466 & 15.9 & 11.5 & 7.90 & 3.10 & 0.990 & 0.071 & 1.33 \\
 & 1.9 & 0.610 & 1.586 & 18.1 & 9.92 & 28.0 & 2.51 & 0.987 & 0.070 & 1.40 \\
 & Mean & & & $15.3 \pm 1.9$ & $12.7 \pm 2.2$ & & $3.92 \pm 1.3$ & $0.995 \pm 0.0074$ & $0.0693 \pm 0.0015$ & $1.29 \pm 0.078$ \\ 
 \\
J1450+5300 & 0.5 & 0.630 & 0.557 & 42.0 & 8.04 & 0.339 & 4.4 & 1.051 & 0.037 & 0.63 \\
 & 1.0 & 0.630 & 0.585 & 49.0 & 6.72 & 2.48 & 3.51 & 1.048 & 0.035 & 0.67 \\
 & 1.5 & 0.630 & 0.502 & 51.6 & 6.04 & 13.4 & 2.25 & 1.036 & 0.038 & 0.70 \\
 & 1.9 & 0.630 & 0.496 & 60.2 & 5.19 & 63.1 & 1.93 & 1.035 & 0.037 & 0.73 \\
 & 1.9 & 0.633 & 0.484 & 60.0 & 5.31 & 63.9 & 1.95 & 1.037 & 0.037 & 0.74 \\
 & Mean & & & $52.6 \pm 6.9$ & $6.26 \pm 1.0$ & & $2.81 \pm 0.98$ & $1.04 \pm 0.0067$ & $0.0368 \pm 0.00098$ & $0.694 \pm 0.040$ \\ 
 \\
J1523+5203 & 0.5 & 0.725 & 0.417 & 182 & 0.560 & 2.20 & 33.0 & 1.188 & 0.0053 & 0.53 \\
 & 1.0 & 0.725 & 0.397 & 228 & 0.485 & 15.8 & 35.7 & 1.189 & 0.0047 & 0.56 \\
 & 1.5 & 0.725 & 0.405 & 265 & 0.438 & 88.3 & 30.0 & 1.187 & 0.0046 & 0.57 \\
 & 1.9 & 0.725 & 0.400 & 307 & 0.399 & 351 & 26.2 & 1.185 & 0.0045 & 0.59 \\
 & 1.9 & 0.715 & 0.372 & 271 & 0.414 & 226 & 23.8 & 1.185 & 0.0040 & 0.56 \\
 & Mean & & & $251 \pm 42$ & $0.459 \pm 0.058$ & & $29.7 \pm 4.3$ & $1.19 \pm 0.0016$ & $0.00462 \pm 0.00042$ & $0.562 \pm 0.019$ \\
\hline
\end{tabular}
\end{table*}


\section{$\beta$ values inferred from X-ray observations of galaxy clusters}\label{beta_from_xray}

\begin{figure}
\includegraphics[scale=0.55, angle=0, trim=0cm 0cm 0cm 0cm]{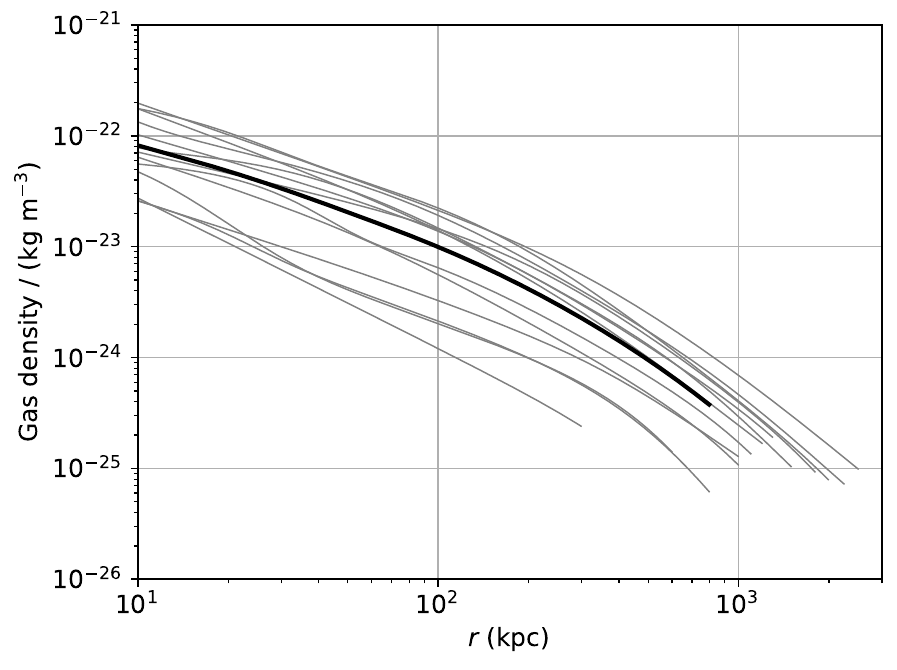}
\caption{Gas density as a function of radius for a sample of 13 nearby, relaxed galaxy clusters derived from \textit{Chandra} observations \citep{vikhlinin2006}. The individual cluster profiles are shown by the thin grey lines. The mean density profile of 11 of the clusters with reliable profile measurements out to $r = 750$~kpc is shown by the thick black line.}
\label{fig:gas_density_profiles}
\end{figure}

\cite{vikhlinin2006} used \textit{Chandra} data to measure the gas density profiles of 13 nearby, relaxed galaxy clusters up to a radius, $r$, of 2.5~Mpc. The gas density profiles shown in figures~3--15 of \citeauthor{vikhlinin2006} are reproduced in Fig.~\ref{fig:gas_density_profiles} using their equation~3. We calculate the mean profile of 11 of the clusters with reliable gas density measurements out to $r = 750$~kpc. The mean cluster profile, shown by the thick black curve, shows continuous steepening with radius.

\begin{figure}
\includegraphics[scale=0.55, angle=0, trim=0cm 0cm 0cm 0cm]{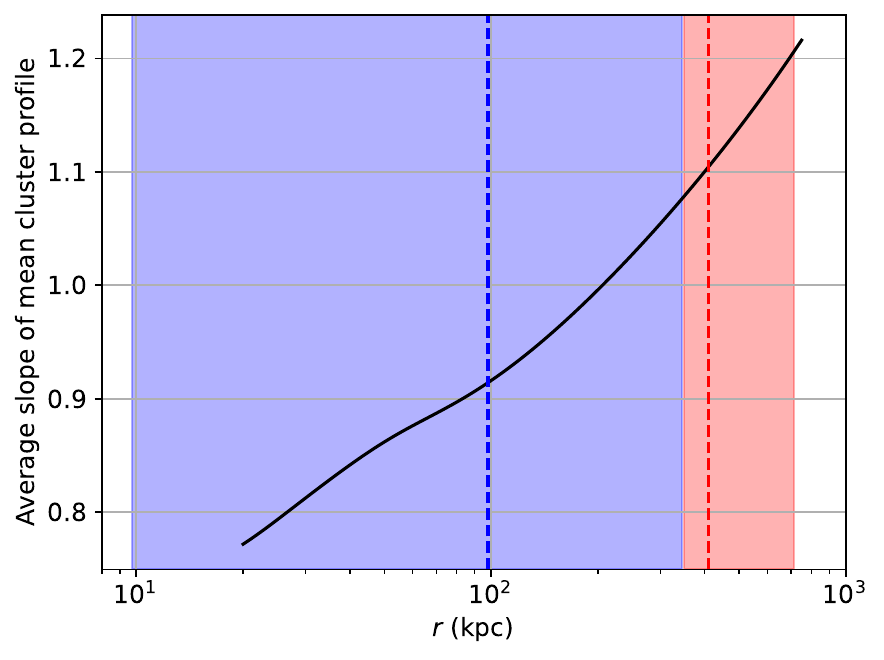}
\caption{Average slope of the mean cluster profile shown in Fig.~\ref{fig:gas_density_profiles} between $a_{0} = 10$~kpc and $r$. The range (median) of $D_{\ell}/2$ for the non-giant RQs in Atlas is shown by the blue shaded region (blue dashed line) and for our GRQs by the red shaded region (red dashed line).}
\label{fig:average_slope_vs_radius}
\end{figure}

We calculate the average slope of the mean cluster profile between $a_{0} = 10$~kpc (the core radius assumed in the dynamical modelling) and $r$, varying $r$ between 20--750~kpc. Fig.~\ref{fig:average_slope_vs_radius} shows that this quantity changes by $\approx 0.4$ between $r = 20-750$~kpc. The non-giant RQs in Atlas and our GRQs have $D_{\ell}$ ranging between $\approx 0.02-0.7$~Mpc and $\approx 0.7-1.43$~Mpc, respectively. The average slope of the mean cluster profile between $a_{0}$ and $D_{\ell}/2$, which is representative of $\beta$, is in the range $\approx 0.8-1.1$ for the non-giant RQs and $\approx 1.1-1.2$ for our GRQs.

\label{lastpage}
\end{document}